\documentclass[10pt,emptycopyrightspace]{ewsn-proc}

\usepackage[inline]{./trackchanges}%
\addeditor{Eric}
\addeditor{Marco}
\addeditor{Qing}

\usepackage{graphicx} 
\usepackage{bmpsize}
\usepackage{amsmath}
\usepackage{algorithmic}
\usepackage{breqn}
\usepackage{array}
\usepackage{balance}
\usepackage{textcomp}
\usepackage{enumitem}
\usepackage{lipsum}
\usepackage{subcaption}
\usepackage{array}
\usepackage{dsfont}
\usepackage{footnote}
\usepackage{color}
\usepackage{xcolor}
\usepackage{comment}
\usepackage{multirow}
\usepackage{tabu}
\usepackage{pgfplots}
\usepackage{xspace}
\usepackage[utf8]{inputenc}
\usepackage[english]{babel}
\usepackage{cite}
\usepackage{soul} 
\usepackage{hyperref}

\addto\extrasenglish{

}

\newcommand{\cardioid}{CardioID\xspace}

\begin{document}

\title{CardioID: Mitigating the Effects of Irregular Cardiac Signals for Biometric Identification}

\numberofauthors{3}
\author{
\alignauthor Weizheng Wang \\
  \affaddr{Delft Univeristy of Technology}
  \email{w.wang-14@tudelft.nl}
\alignauthor Marco Zuniga \\
  \affaddr{Delft Univeristy of Technology}
  \email{m.a.zunigazamalloa@tudelft.nl}
\alignauthor Qing Wang \\
  \affaddr{Delft Univeristy of Technology}
  \email{qing.wang@tudelft.nl}
}




\maketitle



\begin{abstract}
Cardiac patterns are being used to obtain hard-to-forge biometric signatures and have led to high accuracy in state-of-the-art (SoA) identification applications.
However, this performance is obtained under \textit{controlled scenarios} where cardiac signals maintain a relatively uniform pattern, facilitating the identification process.
In this work, we analyze cardiac signals collected in more \textit{realistic (uncontrolled) scenarios} and show that their high signal variability (i.e., irregularity) makes it harder to obtain stable and distinct user features. {Furthermore,} SoA usually fails to identify specific groups of users, rendering existing identification methods futile in uncontrolled scenarios.
To solve these problems, we propose a framework with three novel properties. 
First, we design an adaptive method that achieves stable and distinct features by tailoring the filtering spectrum to each user.
Second, we show that users can have multiple cardiac morphologies, offering us a much bigger pool of cardiac signals and \textit{users}
compared to SoA.
Third, we overcome other distortion effects present in authentication applications with a multi-cluster approach and the Mahalanobis distance.
Our evaluation shows that the average balanced accuracy (BAC) of SoA drops from above 90\% in controlled scenarios to 75\% in uncontrolled ones, while our method maintains an average BAC above 90\% in uncontrolled scenarios.
\end{abstract}


\section{Introduction}
\label{sec:intro}

Biometrics play a fundamental role in human identification.
The popular systems rely on external features, such as fingerprints, iris patterns, and face contours.
These systems have excellent precision but they are vulnerable to attacks: fingerprints can be recreated in latex from touched objects~\cite{kavsaouglu2014novel}; iris patterns can be scanned and emulated~\cite{hern2017samsung}; pictures from the Internet can be used to obtain renditions that can fool face recognition systems~\cite{bhattacharyya2009biometric}.

To overcome the fundamental weakness of \textit{external} features, i.e., the fact that they can be easily captured because they are constantly exposed, researchers are investigating \textit{internal} biometric signals, which are hidden under our skin, and hence, they are hard to obtain and forge.
An approach that is gaining interest is the use of cardiac patterns since they are uniquely defined by the heart, lung and vein structures of an individual. These cardiac patterns can be obtained with a photoplethysmogram (PPG), which measures changes in blood volume via light absorption. PPG signals can be acquired with simple inexpensive sensors that are widely available on wearable devices. For example, one option is to use a pulse oximeter on a finger, which consists of a small LED and a simple photosensor~\cite{karimian2017human}; another option is to place a finger on top of the flashlight and camera in a smartphone~\cite{liu2019cardiocam}. With both types of sensors, researchers have shown that PPG signals can provide between 85\% and 95\% identification accuracy for groups consisting of tens of people~\cite{gu2003firstppg,kavsaouglu2014novel,sarkar2016biometric,liu2019cardiocam}.

\textbf{Challenge.}
The results obtained so far for PPG identification are promising, but they have been obtained mainly under ideal situations: accurate sensors used in controlled environments.  
These {two} factors (sensors and environment) determine how similar cardiac cycles are for the same individual. The higher the similarity of the cardiac cycles, the higher the {identification} accuracy.
We show that when PPG signals are gathered in a more natural (uncontrolled) manner, the cardiac cycles can be highly irregular, significantly decreasing the accuracy of state-of-the-art (SoA) approaches. This uncontrolled environment is common 
in our targeted scenarios: smartphone and door login systems.

\textbf{Our contributions.} Considering the above challenge, we analyse the pernicious effects of irregular cardiac cycles on biometric identification and proposes a novel framework to overcome those effects. In particular, our work provides four main contributions:

\textit{Contribution 1: Morphology Stabilization [\autoref{sec:stabilization}].} The biometric information present in cardiac cycles is restricted to a narrow spectrum of the signal. A key limitation of the SoA approaches is that their filters target the \textit{same} spectrum for \textit{all} individuals. This one-size-fits-all approach leads to either information loss (if the default spectrum is too narrow for a particular individual) or insufficient noise filtering (if the spectrum is too broad). We propose an adaptive filtering technique that fine-tunes the filtering parameters based on the individual cardiac properties. This approach allows us to obtain more stable and distinctive features per user.

\textit{Contribution 2: Morphology Classification [\autoref{sec:feature}].} SoA studies assume that the cardiac pulses of individuals have a \textit{single} dominant morphology (shape). Assuming a single morphology means that several ``non-conforming'' cardiac periods can be unnecessarily discarded, affecting the responsiveness of the system.
{More importantly, we find out that in some cases, the strict SoA assumption of considering a single dominant morphology, leaves out users that rarely have such cardiac morphology, rendering the SoA methods futile for those scenarios.} We show that a \textit{single user} can have \textit{multiple} valid morphologies.
Our ability to consider a wider range of morphologies reduces the amount of time required to test a system, increases the user inclusion to serve more people, and facilitates identifying the rightful individual even when his/her cardiac periods are different from each other.

\textit{Contribution 3}: \textit{Analysis of non-linear effects [\autoref{sec:iden_authen}]}. The SoA utilizes PPG signals to perform two types of biometric applications: identification and authentication. For identification, the SoA uses linear (PCA~\cite{liu2019cardiocam,zhao2013human}, LDA~\cite{yadav2018emotion,sarkar2016biometric}) and non-linear approaches (NN-based~\cite{karimian2017human,spachos2011feasibility}), but there is no analysis determining what approach is better and why. We show that if we tackle the non-linear effects of cardiac cycles at an early stage, both approaches, linear and non-linear, render similar results. For authentication, we identify two main shortcomings in SoA methods: the use of Euclidean distances and the assumption that the features of a subject form a \textit{single} cluster. We propose to leverage \textit{multi}-cluster and Mahalanobis distance to ameliorate non-linear effects.

\textit{Contribution 4}: \textit{Thorough multi-sensor and multi-application evaluations [Section~\ref{sec:eval}]}.  
The evaluation of cardiac signals for biometric applications can be divided into four quadrants: based on the type of sensor (pulse oximeter or camera) and application (identification or authentication).
Most studies evaluate a single quadrant (usually identification with pulse oximeters), \textit{no study has evaluated all four}. Our evaluation includes both types of sensors and applications, and considers cardiac cycles over a wide variability range, from low-variance (controlled) to high-variance (uncontrolled). Overall, our results show that the SoA performs well in controlled scenarios, the average balanced accuracy (BAC) for identification and authentication is above 90\%; but in uncontrolled scenarios, it drops to 75\%. 
Our methods recoup the average BAC above 90\% for uncontrolled scenarios.
\vspace{-1mm}
\section{Preliminaries}
\label{sec:background}

\subsection{PPG basics}
The cardiac cycle represents the continuous change in blood pressure determined by our hearts and blood circulation systems. Given that people have different heart structures in terms of volume, surface shape and motion dynamics~\cite{carlsson2004total,hall2015guyton,lin2017cardiac}, and different tissue thickness and blood vessel distribution~\cite{hall2015guyton}, the cardiac signal has been used to obtain unique biometric signatures~\cite{israel2005ecg,zhao2013human}.
A cardiac cycle can be measured in various manners, the simplest option is to obtain a PPG signal 
by measuring the amount of light absorbed by our body as blood flows through. A PPG signal can be measured with sensors containing inexpensive LEDs and photodiodes, 
or with the flashlight and camera in smartphones. The geometric relations among the various peaks and valleys present in a PPG signal (heights, widths, etc.)~\cite{kavsaouglu2014novel}, or the spectral information in the frequency domain~\cite{karimian2017human}, are optional features used to perform identification. 

\subsection{Applications, morphologies and metrics} 
\label{subsec:principle}

\begin{figure}[t!]
    \centering
    \includegraphics[width=0.9\columnwidth]{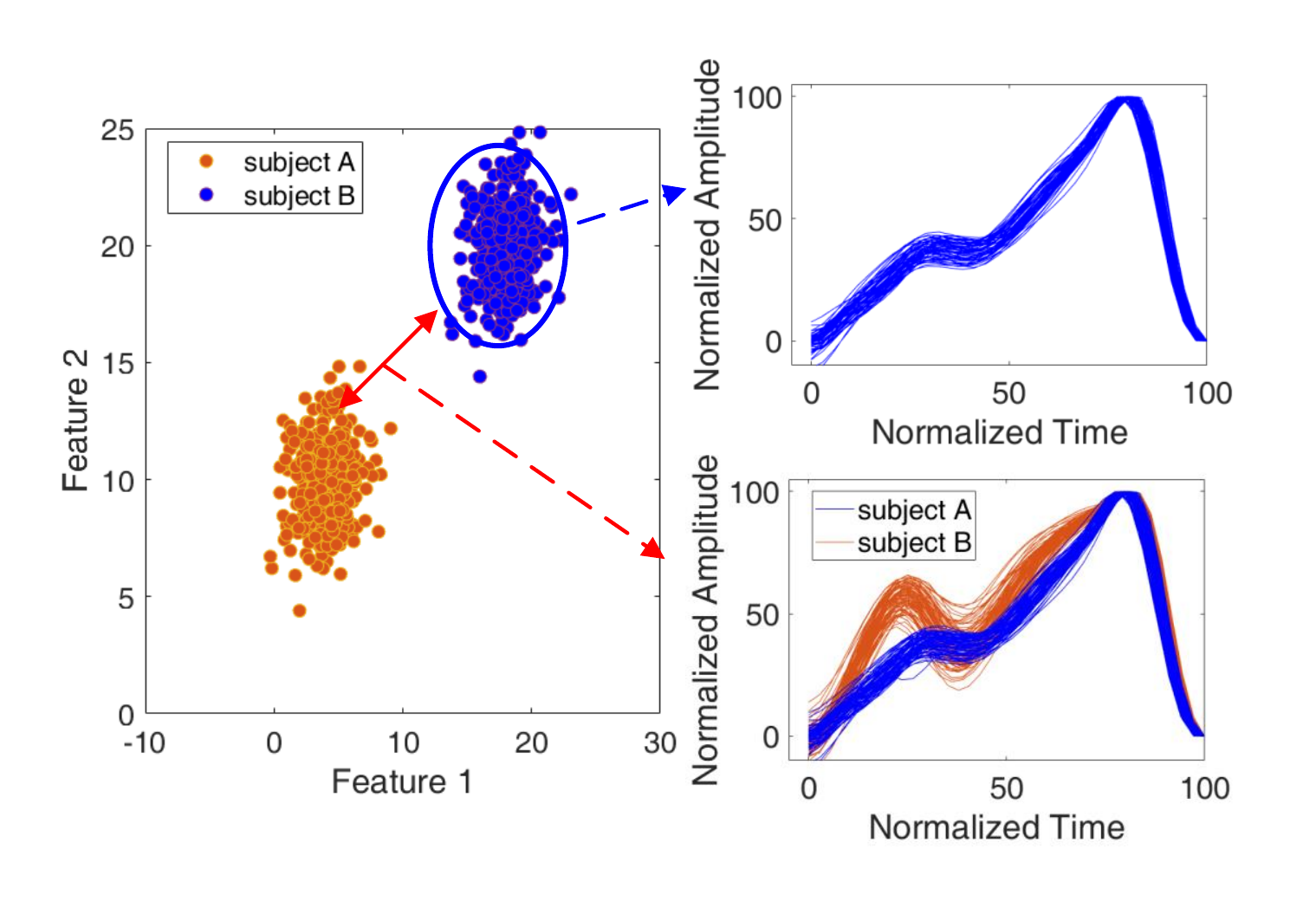}
    \vspace{-6mm}
    \caption{Sample application with \textit{controlled} PPG signals.}
    \vspace{-5mm}
    \label{fig:classification_principle}
\end{figure}

We analyze the performance of two different applications: identification and authentication.
In \textbf{\textit{identification}}, the population size is \textit{known} and the training phase requires gathering data from \textit{all} individuals. The goal is to determine classification boundaries among the \textit{various} subjects.
In \textbf{\textit{authentication}}, the population size is \textit{unknown} and the training phase \textit{only} gathers data from the user of interest. The goal is to determine the best authentication boundary for a \textit{single} subject.

No study in the SoA has tested its methods with both applications: they only focus on one, usually on identification. Our study analyzes both. 
Independently of the target application, achieving high biometric accuracy with PPG signals requires achieving a delicate balance between two competing goals:
\begin{itemize}
\item 
\textbf{\textit{Challenge 1}}: \textit{reduce intra-cluster variance}. We need cardiac cycles that are as homogeneous as possible for the \textbf{same individual}, in order to obtain stable features.
\item 
\textbf{\textit{Challenge 2}}: \textit{increase inter-cluster distance}. We need cardiac
cycles that are as different as possible \textbf{among individuals}, to define clear identification boundaries.
\end{itemize}

\vspace{1mm}
\autoref{fig:classification_principle} shows the PPG signals of two users collected in a controlled manner. Under these favorable circumstances, it is simple to tackle the above challenges and to differentiate the individuals. 

\vspace{1mm}
\textbf{Morphologies.} We use the term \textit{morphology} to refer to the shape of a cardiac cycle, and \textit{stable morphology} to refer to cardiac cycles that have \emph{(i)} the same numbers of peaks and valleys, and \emph{(ii)} a small signal variance. For example, subjects A and B in \autoref{fig:classification_principle} have stable morphologies with two peaks and {three valleys}. In uncontrolled environments, gathering distinct and stable morphologies for each user becomes significantly more complicated.

\textbf{Metrics.} There is no common metric in SoA to measure accuracy. Some studies use the equal error rate (EER) \cite{karimian2017human}, others use F1-score~\cite{kavsaouglu2014novel} or 
BAC~\cite{liu2019cardiocam}. All these metrics 
are derived from true/false positive/negative results. Our results can be presented in any of these metrics. We decide to use BAC because 
{our datasets are imbalance}. BAC is the average of the true positive rate (TPR, sensitivity) and the true negative rate (TNR, specificity):
$BAC=(TPR+TNR)/2 $, where $TPR=True Positive/(True Positive+False Positive)$,
and $TNR=True Negative/(True Negative+False Positive)$.

\begin{figure}[t!]
    \begin{subfigure}{.46\columnwidth}
      \centering
      \includegraphics[width=\linewidth]{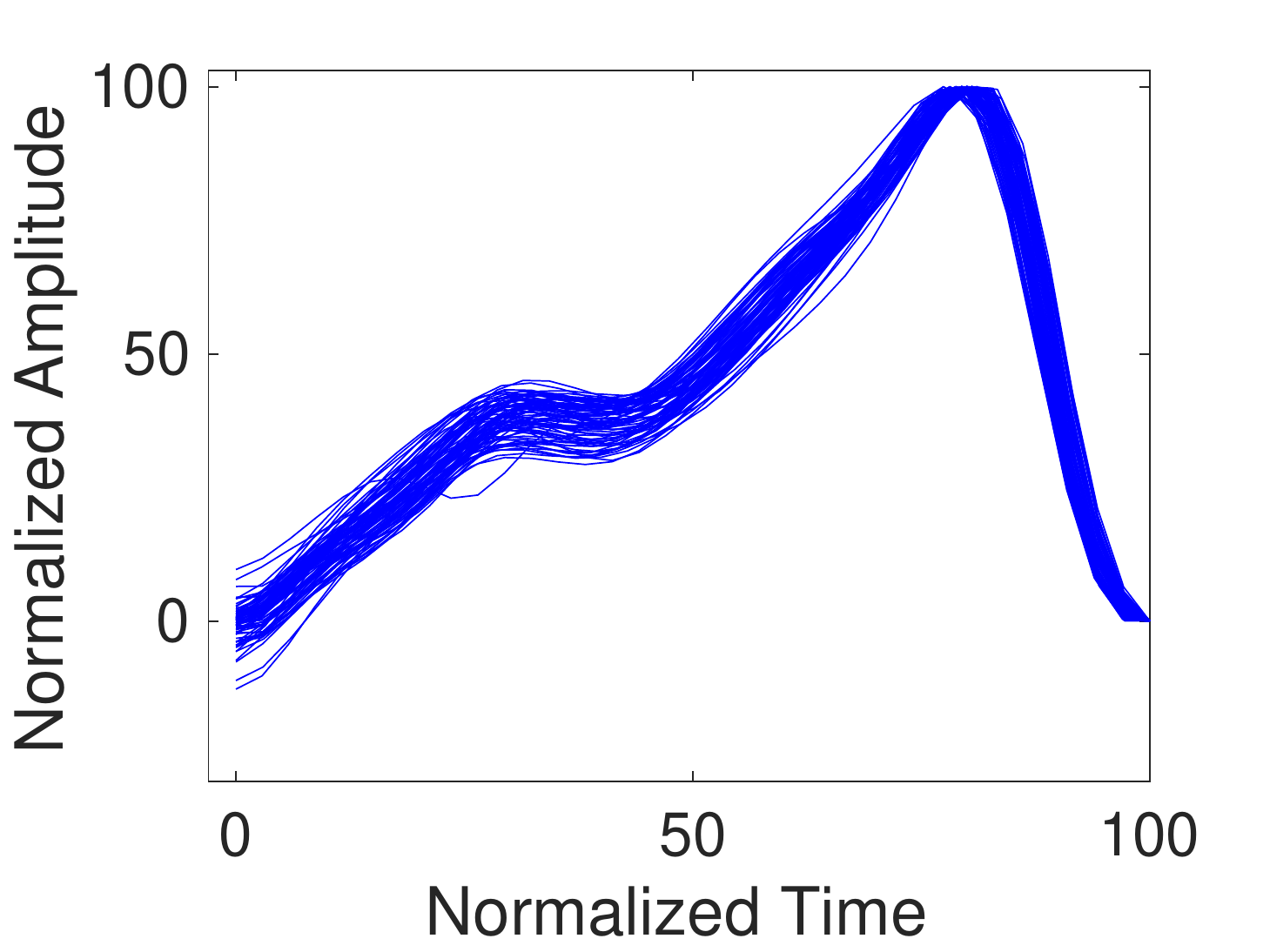}
      \caption{Controlled environment}      
      \label{fig:controlled_ppg}
    \end{subfigure}
    \begin{subfigure}{.46\columnwidth}
      \centering
    \includegraphics[width=\linewidth]{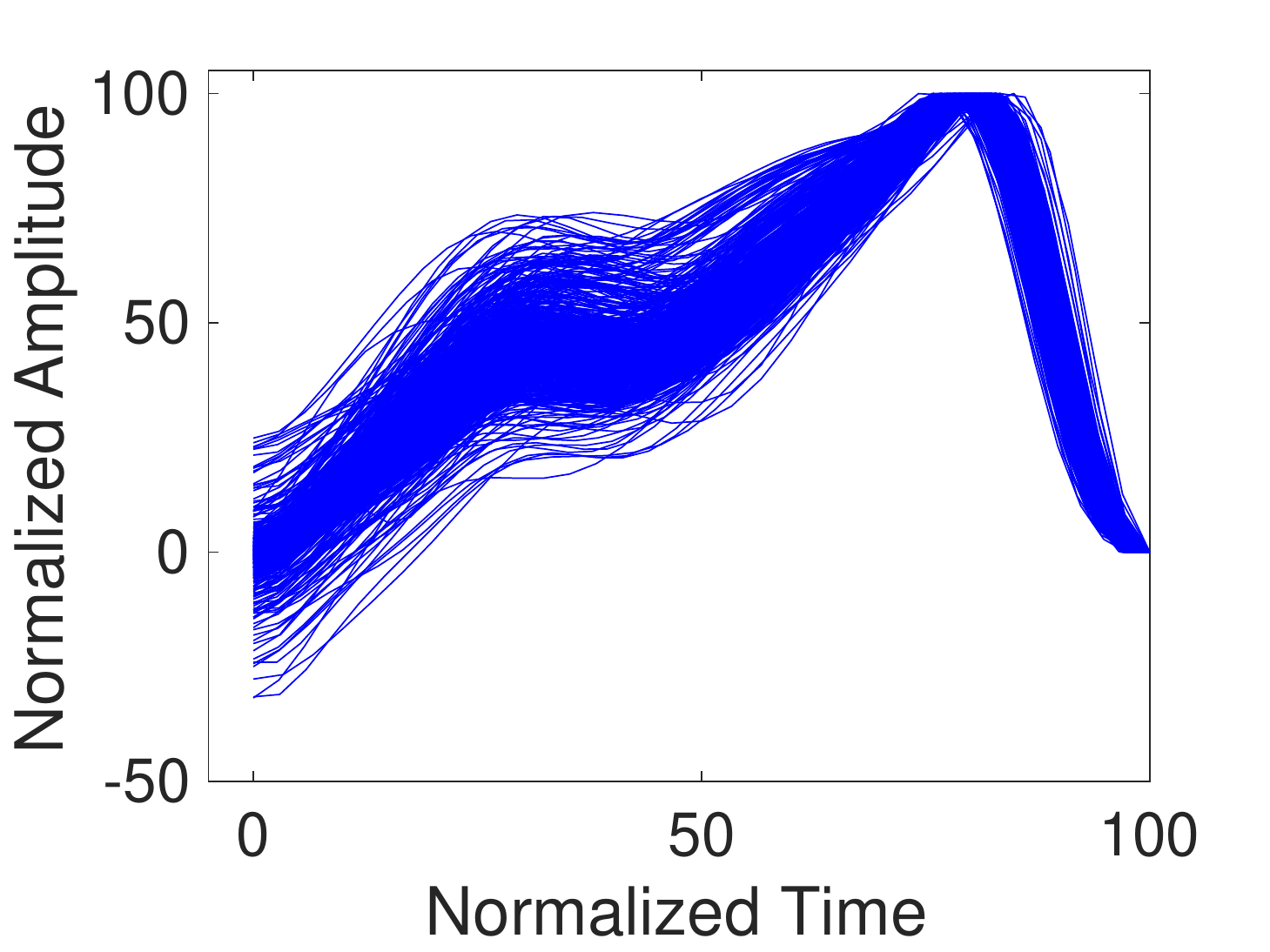} 
      \caption{Uncontrolled environment}
      \label{fig:uncontrolled_ppg}
    \end{subfigure}
    \vspace{-3mm}
    \caption{Cardiac periods collected for the same subject.}
    \vspace{-2mm}
    \label{fig:sample_ppg}
\end{figure}

\begin{table}[!t]
    \centering
    \caption{Performance of SoA.}
    \vspace{-3mm}
    \label{tab:motivation}
    \resizebox{0.95\columnwidth}{!}{
    \renewcommand{\arraystretch}{1}
    \begin{tabular}{|l|c|c|}
        \hline
                                & \textbf{Controlled} & \textbf{Uncontrolled} \\
        \hline
        {Signal Variance}         & 1.83 & 2.96 \\
        \hline
        {BAC for identification \cite{kavsaouglu2014novel}} & 91\% & 72\% \\
        \hline
        {BAC for authentication \cite{liu2019cardiocam}} & 93\% & 69\% \\
        \hline
    \end{tabular}
    \vspace{-5mm}
    }
\end{table}

\subsection{The detrimental effect of irregular cycles}
Multiple PPG studies report a high identification accuracy, ranging from 85\% to almost 100\%, depending on various evaluation parameters and scenarios~\cite{kavsaouglu2014novel,liu2019cardiocam,sarkar2016biometric,gu2003firstppg,spachos2011feasibility,karimian2017human,bonissi2013preliminary}. Most of those studies, however, follow a \textit{well-controlled} data gathering process, which results in limited distortions across cardiac periods, and thus, a good performance. 
The \textit{controlled} process is reflected in two factors: 1) the dataset situation, and 2) the individuals in the dataset. The controlled datasets typically focus on healthy individuals with a narrow age range between 20 and 40 (we will show in \autoref{subsec:datasets}). Thanks to their good cardiac status, their cardiac signals are prone to be stable.
For each individual in the dataset, they take the measurement without minor hand movements and the unconscious pressure change between the fingertip and sensor. So their cardiac signals are even more stable.

In contrast to the \textit{controlled} process, the uncontrolled process is a more realistic daily usage scenario. It covers the age from children to elders and includes minor hand movements and unconscious pressure. 
\autoref{fig:sample_ppg} depicts PPG signals for a single individual collected in controlled and uncontrolled environments. The small variance observed in \autoref{fig:controlled_ppg} is similar to the ones observed in Figure 1 in \cite{sarkar2016biometric} and Figure 3 in \cite{liu2019cardiocam}.\footnote{These studies do not post their PPG data. To infer the variance of their signals, we have to rely on their figures.}
While it is not unreasonable to assume that PPG signals are collected in controlled environments, such assumptions constrain the ubiquitous applicability of PPG-based biometrics.


 
Differences in signal regularity can have a major impact on the performance of SoA methods. \autoref{tab:motivation} shows a preliminary evaluation with four subjects, for whom we collected PPG signals in controlled and uncontrolled environments.
The extraction processes comply with the corresponding papers. In identification evaluation \cite{kavsaouglu2014novel}, there are 758 and 1822 periods for the controlled and uncontrolled environment respectively. In authentication evaluation \cite{liu2019cardiocam}, there are 640 and 1631 periods for the controlled and uncontrolled environment.
The exact description of the SoA methods and the means used to calculate the signal variance are explained in~\autoref{subsec:baselines} ~\autoref{subsec:datasets}, respectively. For now, \emph{the important takeaway is that when the SoA is tested with controlled data, the performance is high, as reported in the original studies; but when tested with highly variable signals, the accuracy drops significantly.}

\section{Morphology {Stabilization}} \label{sec:stabilization}

\begin{figure}[t!]
 \centering
  \includegraphics[width=\columnwidth]{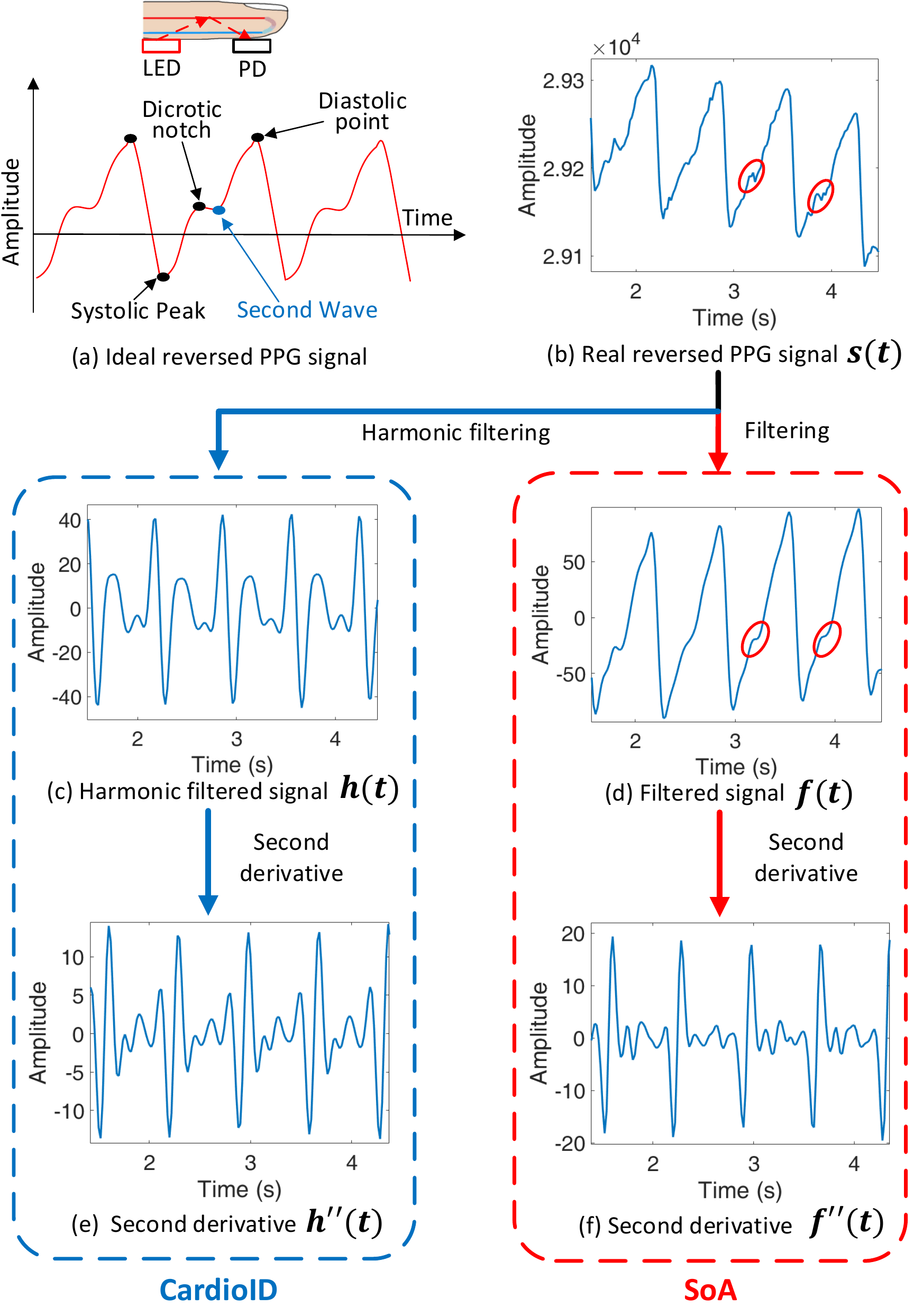}
  \vspace{-7mm}
  \caption{Morphology stabilization.}
  \vspace{-5mm}
  \label{fig:morpholgy_stabilization}
\end{figure}

\begin{figure*}[t!]
  \centering
  \begin{minipage}{.72\textwidth}
    \centering
    \begin{subfigure}{0.33\textwidth}
        \vspace{-3mm}
        \includegraphics[width=\linewidth]{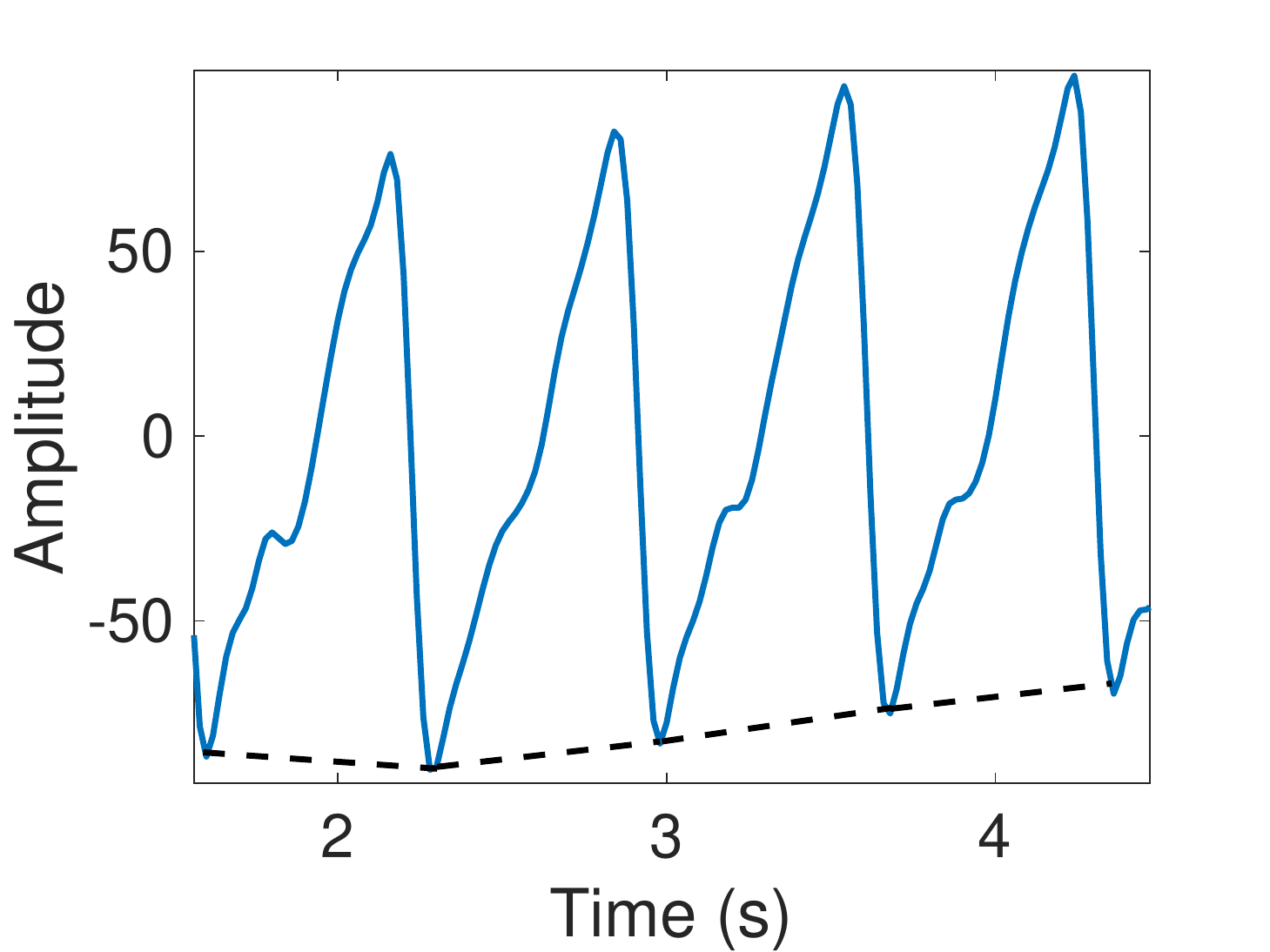}
        \vspace{-5mm}
        \caption{$f(t)$ in a sliding window}
        \label{fig:sa_1}
    \end{subfigure}\hfil
    \begin{subfigure}{0.33\textwidth}
        \includegraphics[width=\linewidth]{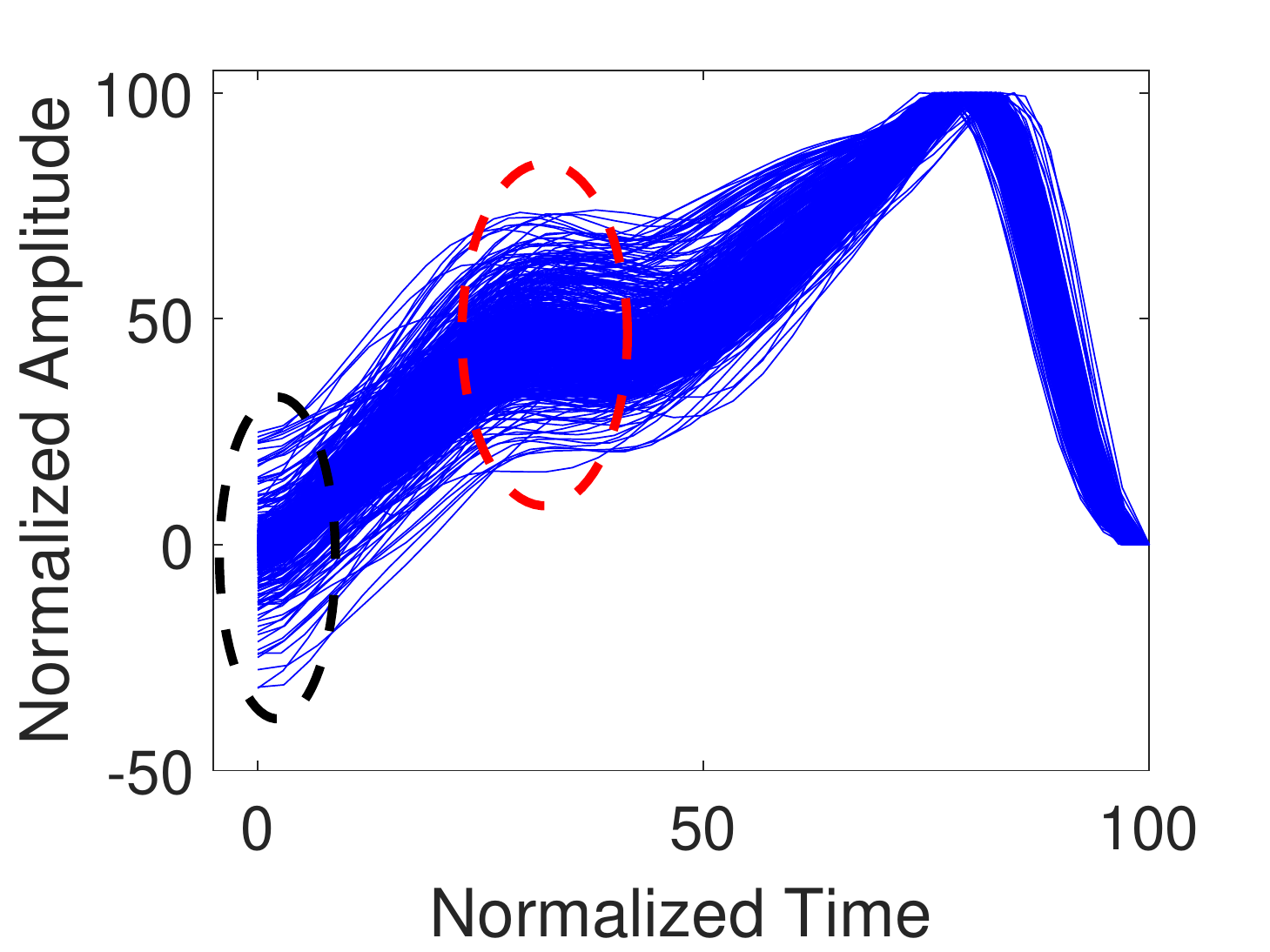}
        \vspace{-5mm}
        \caption{Overlapping all periods \protect\\of $f(t)$}
        \label{fig:sa_2}
    \end{subfigure}\hfil
    \begin{subfigure}{0.33\textwidth}
        \includegraphics[width=\linewidth]{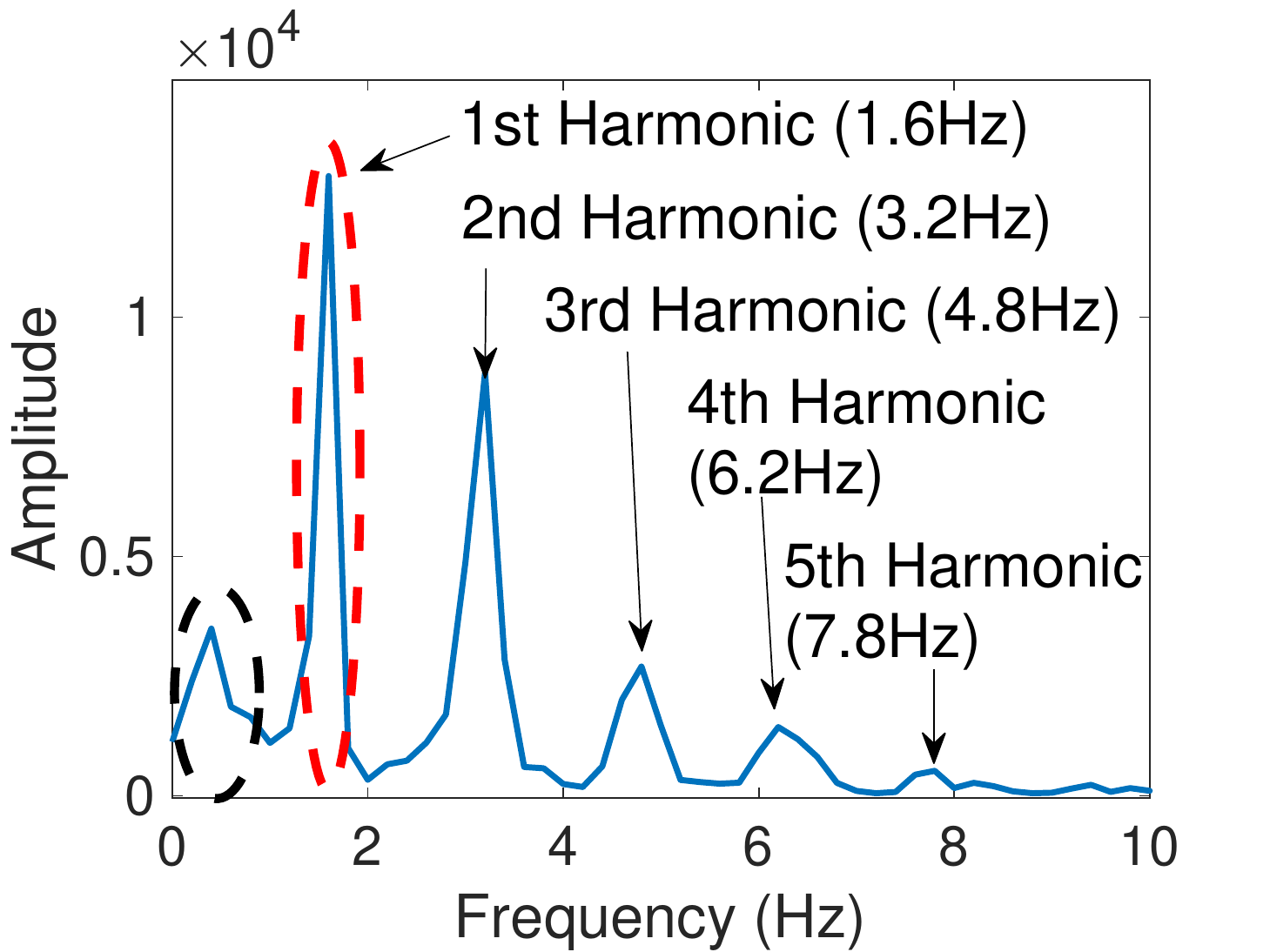}
        \vspace{-5mm}
        \caption{Spectral domain of $f(t)$ in \protect\\the sliding window}
        \label{fig:sa_3}
    \end{subfigure}
    \medskip
    \begin{subfigure}{0.33\textwidth}
      \vspace{-3mm}
      \includegraphics[width=\linewidth]{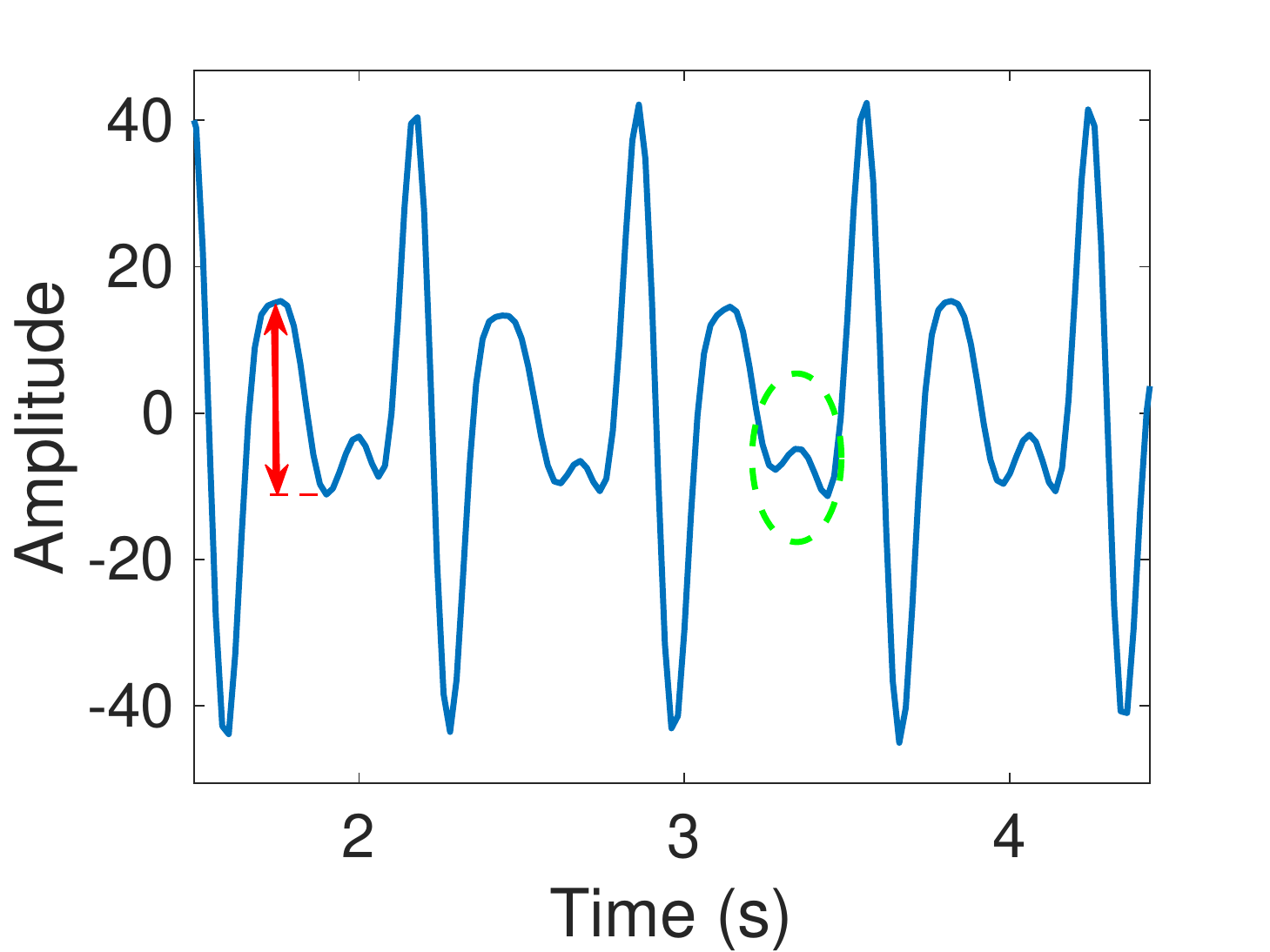}
      \vspace{-5mm}
      \caption{$h(t)$ in the sliding window}
      \label{fig:sa_4}
    \end{subfigure}\hfil
    \begin{subfigure}{0.33\textwidth}
      \includegraphics[width=\linewidth]{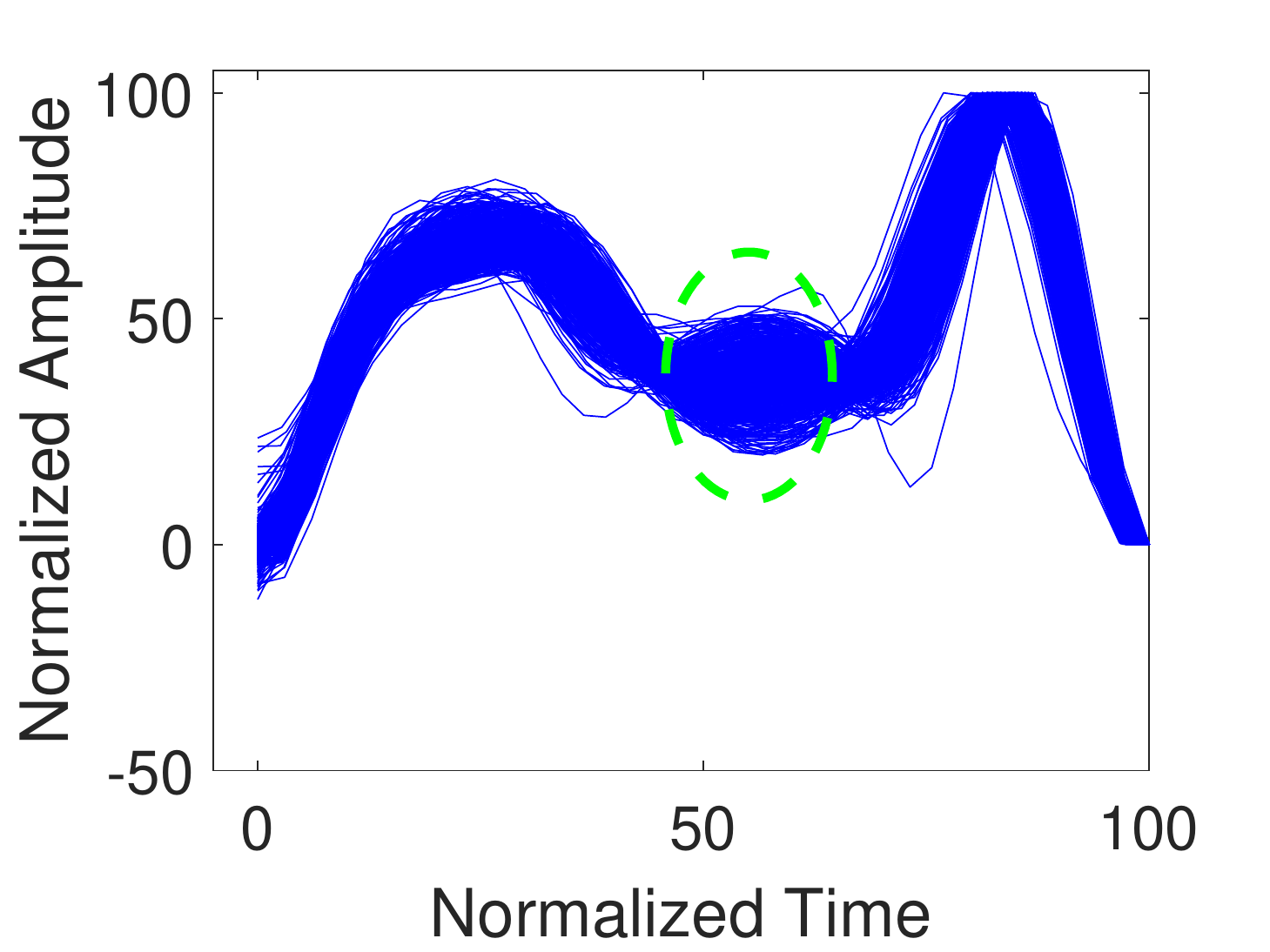}
      \vspace{-5mm}
      \caption{Overlapping all periods \protect\\ of $h(t)$}
      \label{fig:sa_5}
    \end{subfigure}\hfil
    \begin{subfigure}{0.33\textwidth}
      \includegraphics[width=\linewidth]{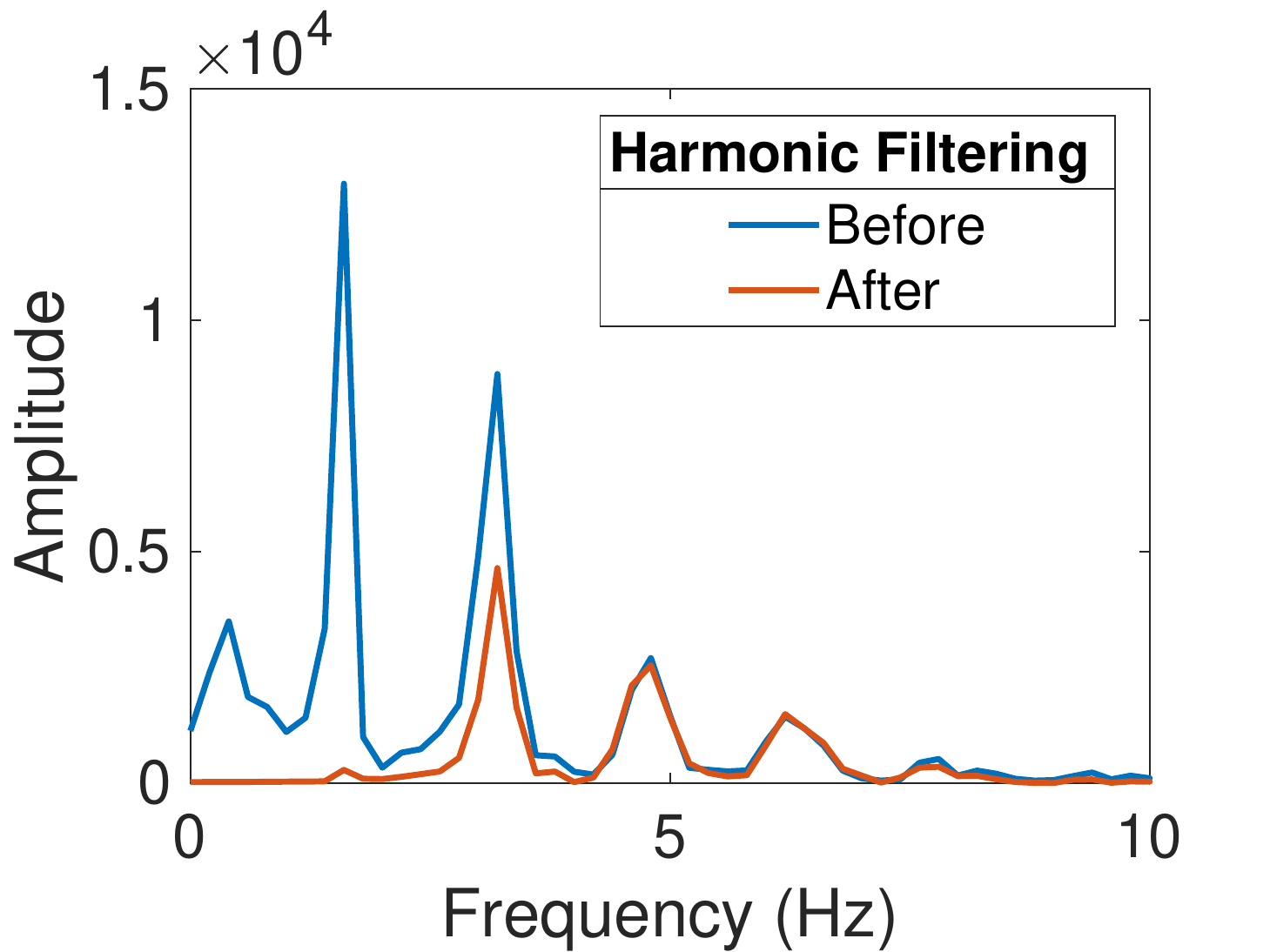}
      \vspace{-5mm}
      \caption{Spectral domain of $h(t)$ in \protect\\the sliding window}
      \label{fig:sa_6}
    \end{subfigure}
    \vspace{-5mm}
    \caption{Frequency analysis to determine the lower cut-off frequency $f_l$.}
    \vspace{-5mm}
    \label{fig:situation_analyze}
  \end{minipage}
  \hfill
  \begin{minipage}{.23\textwidth}
  \vspace{4mm}
  \centering
    \includegraphics[width=\linewidth]{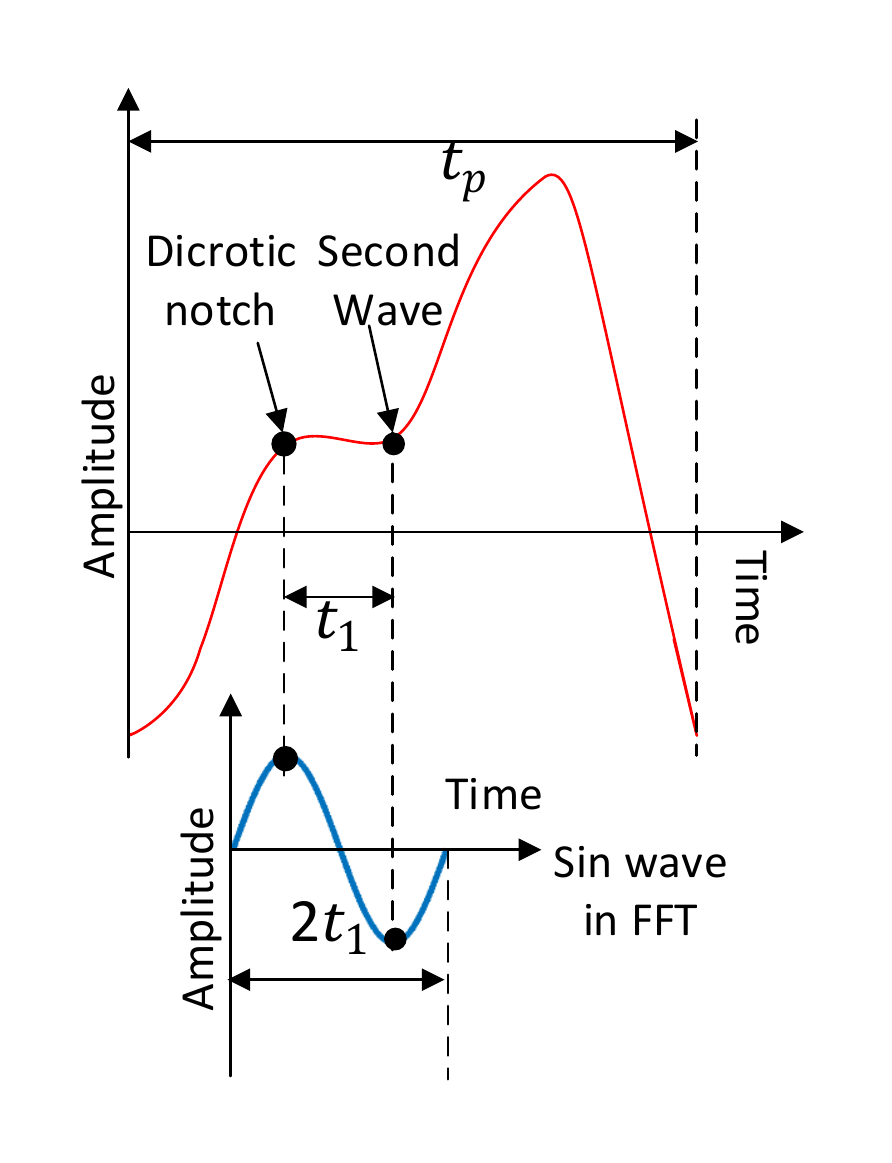}
    \vspace{-6mm}
    \caption{Frequency analysis to determine the upper cut-off frequency $f_h$.}
    \vspace{-15mm}
    \label{fig:upper_bound}
  \end{minipage}
\end{figure*}

A major shortcoming of the SoA is to use the \textit{same spectrum} to filter the PPG signals of \textit{all subjects}. 
In this section, we propose a novel adaptive filtering method. 
\autoref{fig:morpholgy_stabilization} depicts a macro view of our approach and its relation with the SoA. First, we describe the methods we borrow from the SoA (\autoref{subsec:filtering_soa}), and then, we describe their limitation and present our contributions (\autoref{subsec:filtering_own}).

\subsection{From SoA: Basic filtering and derivatives}
\label{subsec:filtering_soa}
\autoref{fig:morpholgy_stabilization}a depicts an ideal PPG signal. The biometric signature of an individual is captured by four fiducial points: diastolic (highest valley), systolic (lowest peak), the dicrotic notch (which form a small peak in the middle of the period) and second wave. \autoref{fig:morpholgy_stabilization}b shows a raw PPG signal $s(t)$, which has two undesirable properties. First, a significant amount of noise distorts the location and intensity of the fiducial points, and in some cases, the noise level can be so high to erase the second wave and dicrotic notch completely, affecting the system's accuracy severely. Second, even in the ideal case, when all fiducial points are present, the signal's morphology is too simple and generic. Given that features are obtained based on the relative duration, heights, and slopes between fiducial points, the limited number of fiducial points limits the number of features. To overcome these effects, the SoA proposes a basic filtering step and the use of the second derivative of the PPG signal.

\textbf{Filtering.}
To mitigate the noise in PPG signals, SoA has identified the spectrum over which cardiac information is contained. 
For biometric purposes, the lowest meaningful frequency of a PPG signal is the heart rate. Considering that athletes can have heart rates as low as 0.5\,Hz \cite{Athlete2018heartrate}, the lower cut-off frequency $f_l$ is usually set to that value. Regarding the upper cut-off frequency $f_h$, according to \cite{choi2017PPG}, sampling frequencies above 25\,Hz do not provide any extra information, hence, $f_h$ can be set to 12.5\,Hz (due to the Nyquist-Shannon sampling theorem). Some studies use other filtering bands \cite{kavsaouglu2014novel,liu2019cardiocam}, but the overall filtering process is similar. Figure~\ref{fig:morpholgy_stabilization}d shows a PPG signal $f(t)$ after being filtered with a second-order Butterworth bandpass filter with bandwidth 0.5-12.5\,Hz \cite{butterworth1930theory}.

\textbf{Derivatives.}
Filtering alleviates noise, but it also eliminates valuable information.
For instance, the raw PPG signal $s(t)$ in~\autoref{fig:morpholgy_stabilization}b contains faint but detectable second waves (red circles). After filtering, however, those fiducial points no longer exist (corresponding red circles in~\autoref{fig:morpholgy_stabilization}d).
To overcome this issue, researchers obtain features not only from $f(t)$ but also from its second derivative $f''(t)$ \cite{kavsaouglu2014novel}. \autoref{fig:morpholgy_stabilization}f depicts the second derivative, which exhibits more fiducial points than $f(t)$.

\subsection{Contribution: Harmonic filtering} 
\label{subsec:filtering_own}

We also use the filtering and derivative stages, but we do not utilize the same parameters for all users. 
We propose a harmonic filtering phase that adapts its parameters to every user. This process allows us to obtain more stable morphologies for every user (Challenge 1) and distinct fiducial points among users (Challenge 2). Considering our harmonic filtering depends on the subjects' heartrates that can change over time, our system proceeds signals based on a 5-second sliding window with a 1-second stride, to track the change of subjects' heartrates.

\vspace{-1mm}
\subsubsection{Determining the lower cut-off frequency $f_l$} The SoA usually uses a lower cut-off frequency that is too low, which increases signal variance and makes it hard to identify the most vulnerable fiducial points (second wave and dicrotic notch). \autoref{fig:situation_analyze}a shows the filtered signal $f(t)$ using SoA methods and~\autoref{fig:situation_analyze}b shows the overlapping cardiac cycles using the end point of periods as an alignment anchor. We can observe a large variance in the starting points (black ellipsoid in~\autoref{fig:situation_analyze}b) and significant instability in the dicrotic notch (red ellipsoid in~\autoref{fig:situation_analyze}b).

Thus, \emph{the fundamental question is how high should $f_l$ be?} To obtain this optimal value, we analyze $f(t)$ in the spectral domain in~\autoref{fig:situation_analyze}c. Our analysis leads to two important insights. First, the wide variance occurs because an $f_l=0.5$\,Hz does not filter important dynamics such as heart rate variability, the effect of respiration (slow changing frequency component) and subtle unconscious pressure changes on the fingertip, which are common phenomena in uncontrolled scenarios. 
Those dynamics generate a fluctuating envelope in the time domain (black dashed line in \autoref{fig:situation_analyze}a), which causes the height differences between the starting and end points in periods. Considering the end points are the alignment anchors, those height differences among periods will lead to a significant variance in the starting points.
Second, an $f_l=0.5$\,Hz obscures the dicrotic notch. The energy of PPG is concentrated around the harmonics of the heartbeat, in particular the first harmonic (red ellipsoid in~\autoref{fig:situation_analyze}c). The SoA does not filter the first harmonic for its period as a feature, which is good, but the spectral energy of the heart rate overwhelms the second wave and dicrotic notch, which are the most vulnerable fiducial points. Furthermore, it is unnecessary to keep the first harmonic to obtain the heart rate because it is still contained on the other harmonics. 

Our analysis indicates that to ameliorate the dampening effects of the heart rate period, we need to filter out the first harmonic. We noticed, however, that for some subjects the second harmonic is as high (and as dampening) as the first harmonic and should be attenuated too. Therefore, denoting the frequency of the first harmonic as $f_{1h}$, we set $f_l=2f_{1h}$.

\subsubsection{Determining the upper cut-off frequency $f_h$} High frequency noise modifies the location of fiducial points, which in turn, affects the stability of features and the overall performance of the system. Depending on the individual, a $f_h = 12.5$\,Hz may be too high. For example, in~\autoref{fig:situation_analyze}c the spectral energy is almost negligible beyond 10\,Hz. Considering this situation, how low should $f_h$ be?  

As stated earlier, it is central to preserve the most vulnerable fiducial points on PPG signals (second wave and dicrotic notch). We use Figure~\ref{fig:upper_bound}, which zooms into those two vulnerable points, to illustrate the derivation of $f_h$. Denoting $t_1$ as the duration between the second wave and the dicrotic notch, the sine wave in the FFT containing the spectral energy of these points has a period of $2t_{1}$, which means that $f_h$ must be higher than $1/{2t_{1}}$, else those two fiducial points would be filtered out. Now, denoting $t_p$ as the period of a cardiac cycle, we observed empirically that $2t_{1} > t_p/5$, and consequently, in the frequency domain $1/2t_{1} < 5/t_p$. Finally, considering that $5/t_p$ represents the fifth harmonic of the heartrate, we set $f_h={5.5}{f_{1h}}$ to preserve all fiducial points while removing high frequency noise. The negligible frequency components beyond the fifth harmonic in \autoref{fig:situation_analyze}c prove the correctness of our analysis.

\subsubsection{Adaptive filtering} The frequency response of our filter is solely based on the first harmonic, $2f_{1h}$ to $5.5f_{1h}$, which is simple to obtain from the signals. More importantly, our approach is based on the subject's heartbeat instead of fixed parameters, allowing us to perform accurate adaptive filtering per subject. \autoref{fig:situation_analyze}d, \autoref{fig:situation_analyze}e and \autoref{fig:situation_analyze}f show the signals filtered with our method, their overlapping cycles and spectral domains. We can observe that, compared to the filtered signal $f(t)$~in~SoA, $h(t)$ has three advantages: \emph{(i)} the signal variance is much lower throughout the entire cycle, \autoref{fig:situation_analyze}e; \emph{(ii)} the difference between the second wave and dicrotic notch is accentuated significantly, {red arrow} in~\autoref{fig:situation_analyze}d; and \emph{(iii)} our method exposes another fiducial point, green ellipsoid in Figure~\ref{fig:situation_analyze}d, which we can exploit to obtain more features as explained next.

\subsubsection{Derivatives} 
\begin{figure}[t!]
    \begin{subfigure}{.47\columnwidth}
      \centering
      \includegraphics[width=\linewidth]{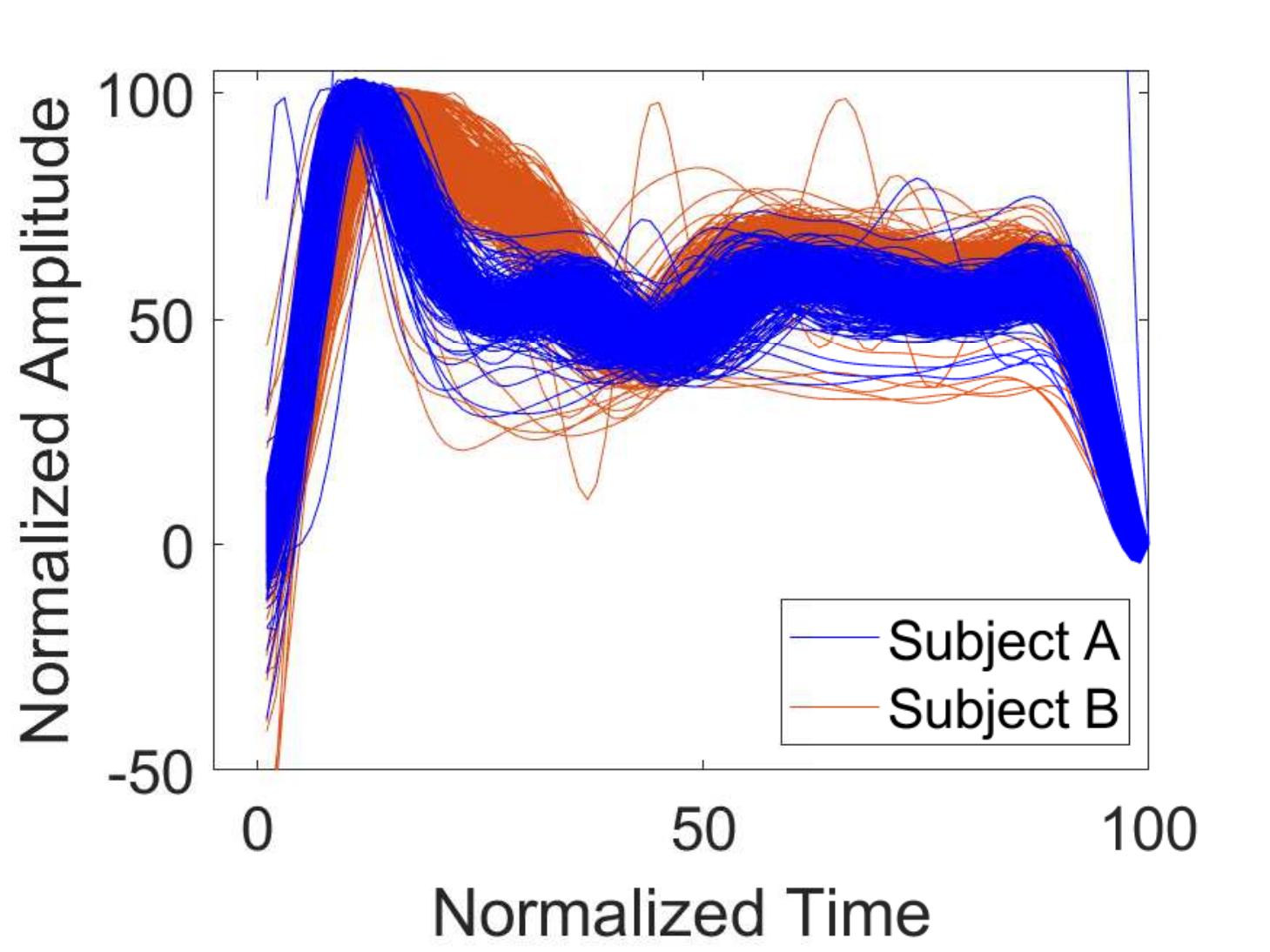} 
      \caption{SoA: $f''(t)$}
      \label{fig:HR_superposition}
    \end{subfigure}
    \begin{subfigure}{.47\columnwidth}
      \centering
      \includegraphics[width=\linewidth]{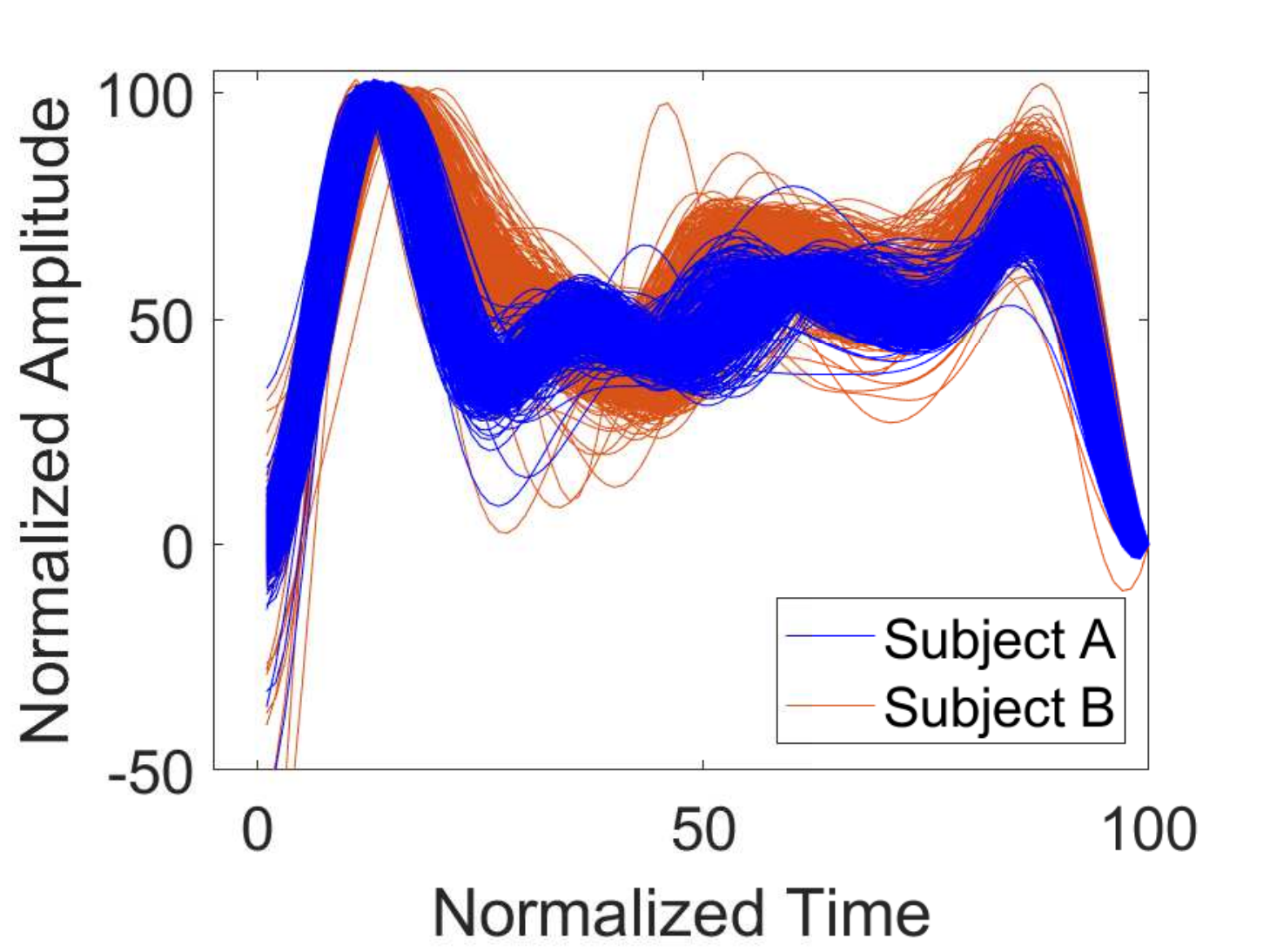}
      \caption{CardioID: $h''(t)$}
      \label{fig:SD_superposition}
    \end{subfigure}
    \vspace{-2mm} 
    \caption{Overlapping periods with uncontrolled data.}
    \vspace{-5mm} 
    \label{fig:SD_improvement}
\end{figure}

As described earlier, the SoA uses derivatives to accentuate the presence of fiducial points. We borrow that idea to obtain the second derivative of our harmonic signal $h(t)$. \autoref{fig:SD_improvement} plots overlapping cycles for $f''(t)$ and $h''(t)$ for two sample subjects with uncontrolled data. 
Our second derivative $h''(t)$ has two important advantages compared to $f''(t)$. First, even though $f''(t)$ is more stable than $f(t)$ because the derivative removes offsets, $h''(t)$ is still less variable because it inherits the stability of $h(t)$. The variance of $f''(t)$ for subjects A and B are {2.8 and 3.0}, respectively, while for $h''(t)$ are 
2.2 and 2.8. This lower variability helps to tackle \textit{Challenge 1}. 
Second, thanks to the tailored cut-off frequencies of our adaptive filter, $h''(t)$ can exploit the specificity of $h(t)$ to obtain more distinctive morphologies for different users, tackling \textit{Challenge 2}. Compared to $f''(t)$, the fiducial points of $h''(t)$  are more distinctive and conspicuous across the \textit{entire} time domain.
Furthermore, subject A (blue) in \autoref{fig:SD_superposition} shows that the second derivative disentangles the `knot' caused by the new fiducial point captured by the green ellipsoid in~\autoref{fig:situation_analyze}e.

\textbf{Summary.} Overall, our approach also follows the two basic steps of the SoA, filtering and second derivatives, but using a novel filtering method leads to a more stable morphology for each user (Challenge 1) and more distinctive morphologies for different users (Challenge 2). The only input parameter required by our filter is the first harmonic (heart rate period), which can be easily obtained from any PPG signal. SoA obtains its features from $f(t)$ and $f''(t)$, and we obtain them from $h(t)$ and $h''(t)$. An exact description of what features we use is provided in the next section.
\section{Morphology Classification}\label{sec:feature}


Existing studies share a common underlying assumption: all cardiac signals have a single dominant morphology.
That, however, is not necessarily the case. We show that a single user can have multiple \textit{valid} morphologies. Without this insight, a system would need to either discard periods that do not conform to a pre-defined morphology (introducing latency), or consider all periods with different morphologies, but at the risk of obtaining widely different features for the same user (reducing accuracy). 

In this section, we first show that cardiac periods can have multiple morphologies, then we describe the features derived from those various morphologies.

\subsection{Multiple morphologies}
\label{subsec:morphology_classification}

\begin{figure}[t]
     \centering
     \vspace{-3mm}
      \begin{subfigure}{.32\linewidth}
          \centering
          \includegraphics[width=\linewidth]{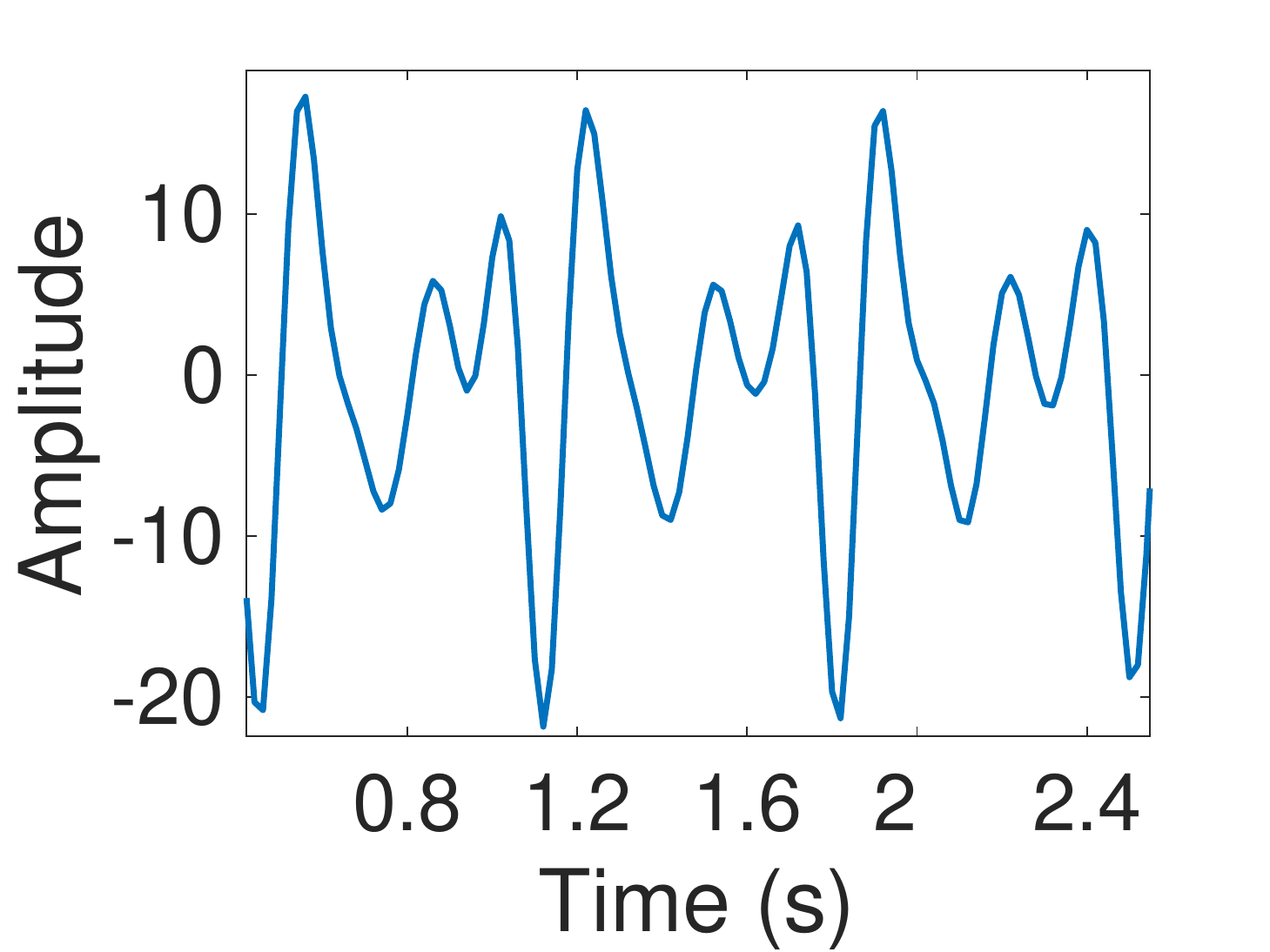}
          \vspace{-5mm} 
          \caption{$h''_1(t)$: 3 peaks, 4 valleys}
        \label{fig:M1}
        \end{subfigure}
        \begin{subfigure}{.32\linewidth}
          \centering
          \includegraphics[width=\linewidth]{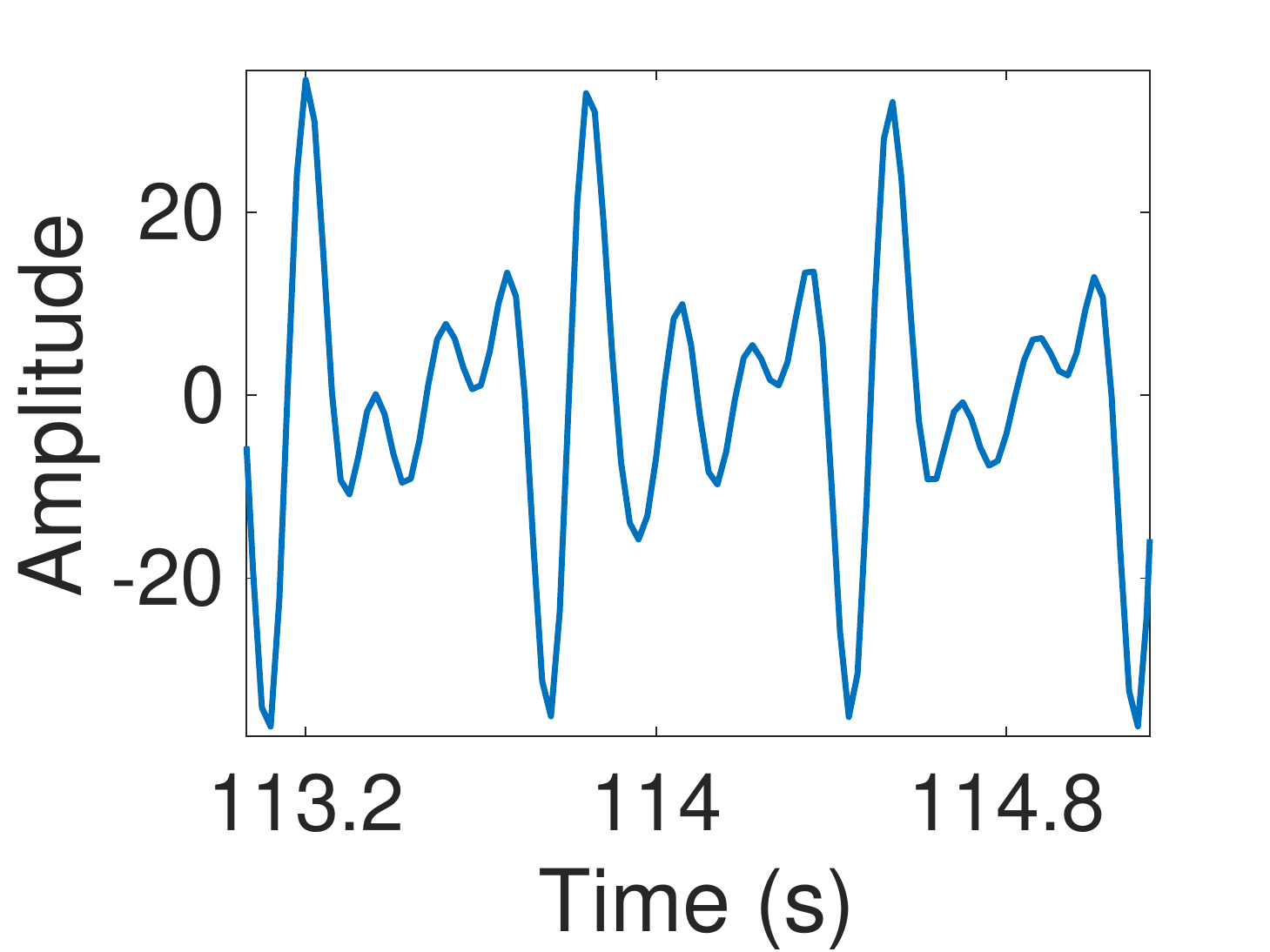}
          \vspace{-5mm} 
          \caption{$h''_2(t)$: 4 peaks, 5 valleys}
        \label{fig:M2}
        \end{subfigure}
        \begin{subfigure}{.32\linewidth}
          \centering
          \includegraphics[width=\linewidth]{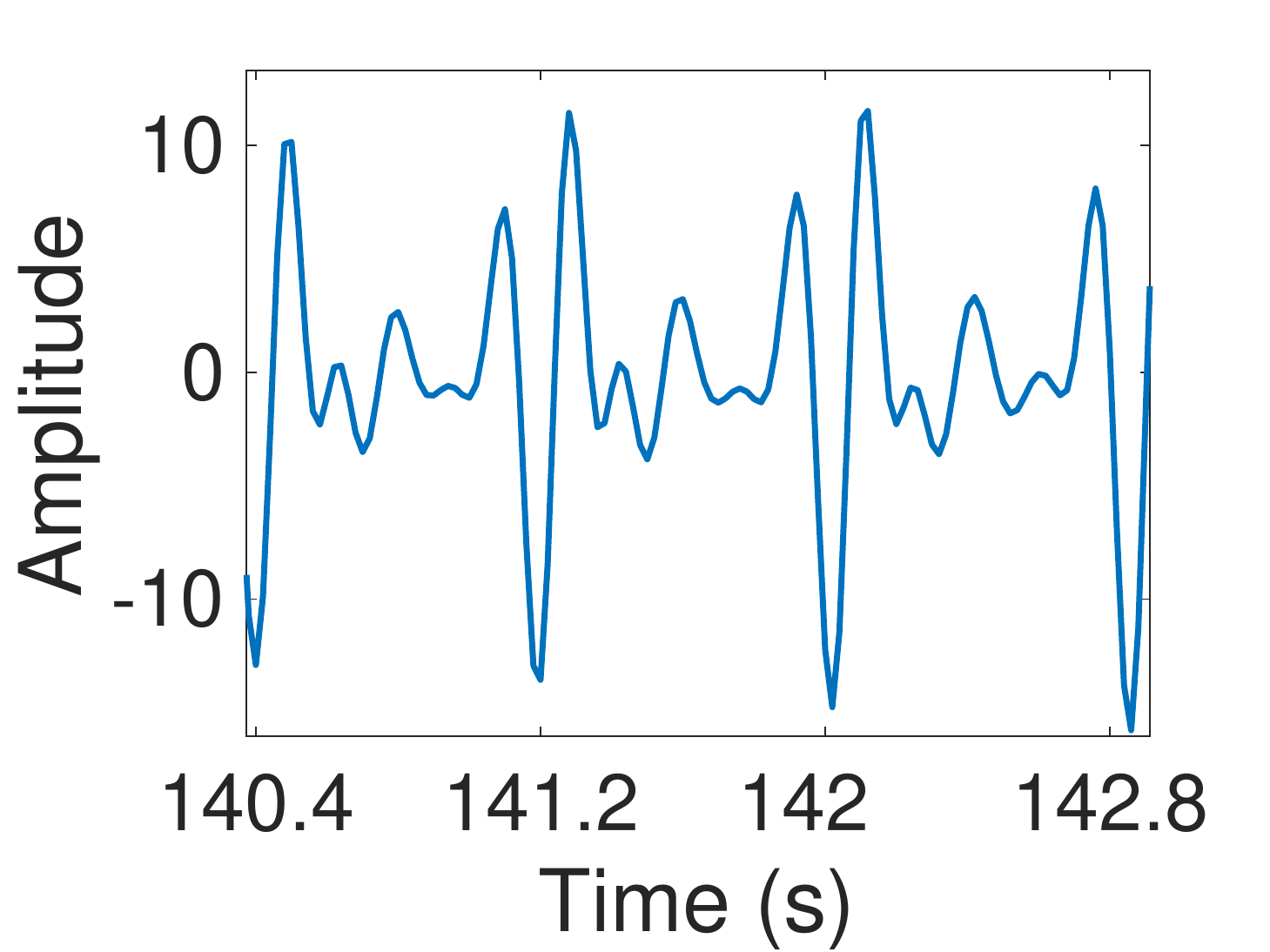}
          \vspace{-5mm} 
          \caption{$h''_3(t)$: 5 peaks, 6 valleys}
        \label{fig:M3}
        \end{subfigure}
        \vspace{-2mm} 
        \caption{Dominant morphologies for $h''(t)$.}
        \vspace{-5mm} 
        \label{fig:dominated_morphologies}
\end{figure}

\begin{figure}[t]
    \centering
    \begin{subfigure}{.49\columnwidth}
        \centering
        \includegraphics[width=\linewidth]{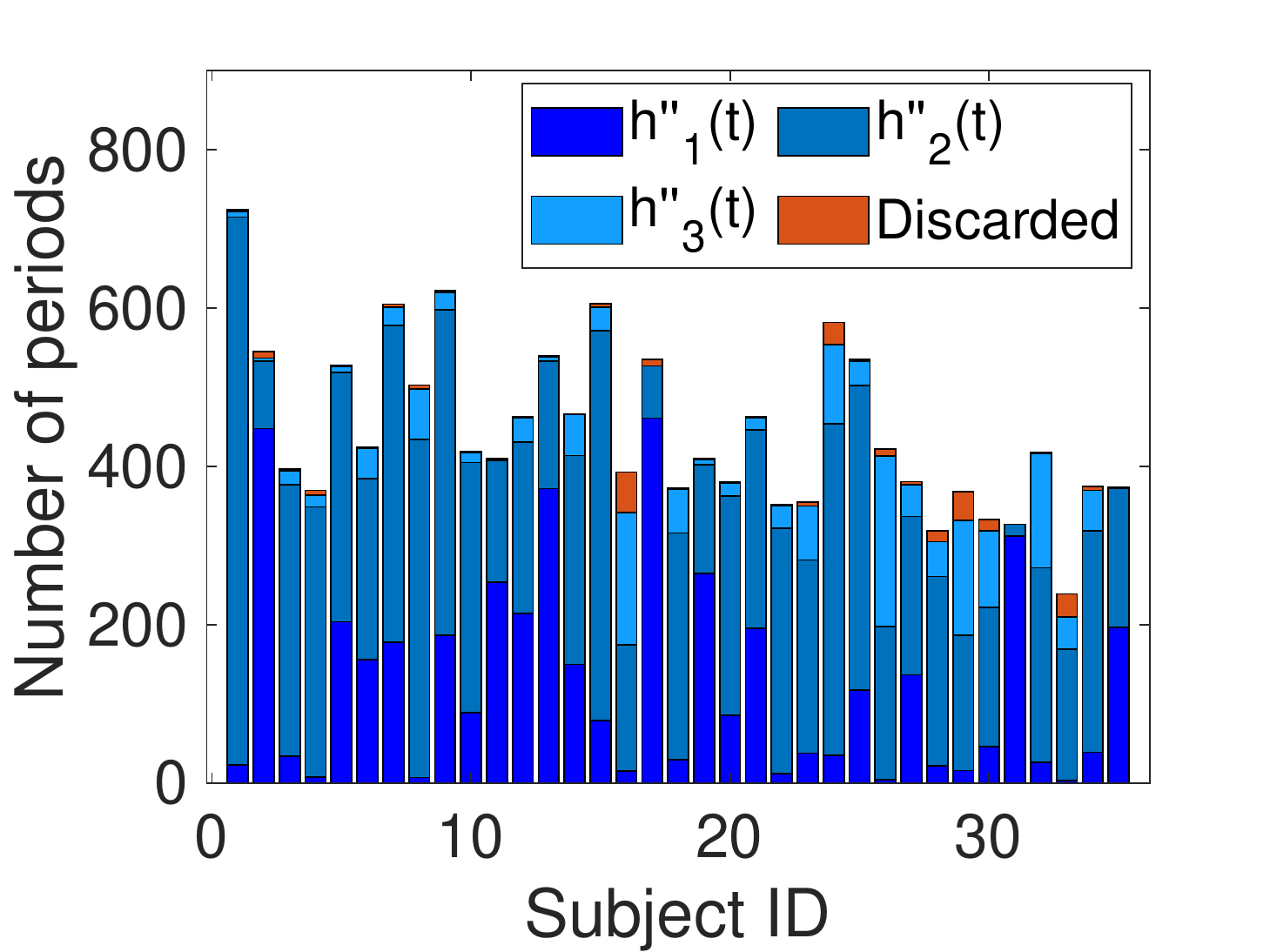}
        \vspace{-5mm}
        \caption{Pulse oximeter dataset}
        \label{fig:periods_pulse_oximeter}
    \end{subfigure}
    \hfill
    \begin{subfigure}{.49\columnwidth}
        \centering
        \includegraphics[width=\linewidth]{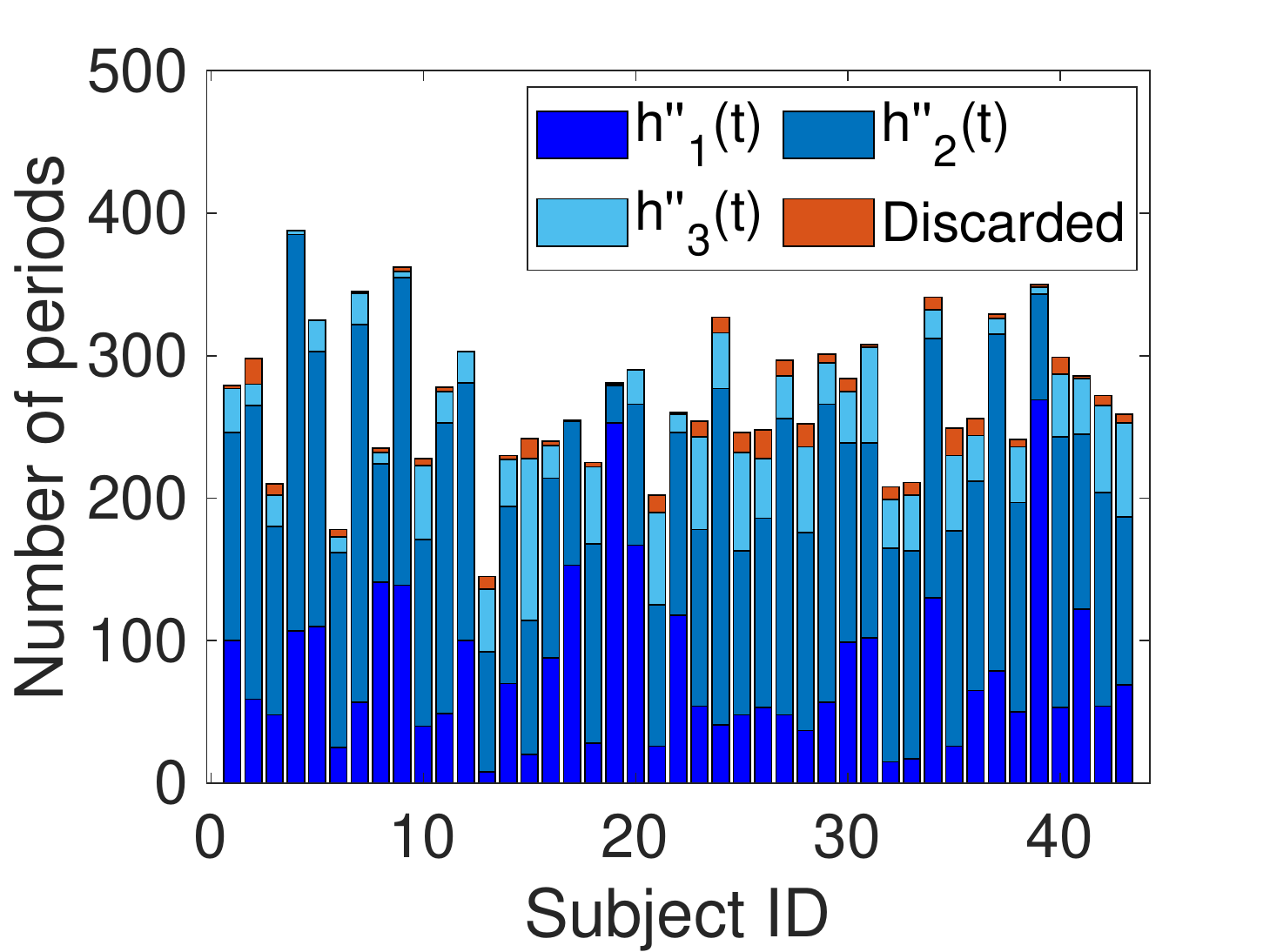}
        \vspace{-5mm}
        \caption{Camera dataset}
        \label{fig:periods_camera}
    \end{subfigure}
    \vspace{-3mm}
    \caption{Frequency of occurrence of morphologies.}
    \vspace{-6mm}
    \label{fig:frequency_morphologies}
\end{figure}

Currently, all studies using fiducial points assume a single macro morphology for all subjects. That is a valid approach in controlled scenarios, but in uncontrolled scenarios various factors can cause the appearance of multiple morphologies: unintended fingertip pressure~\cite{ppgpressure2020}, significant differences in the cardiac profiles of subjects, etc.
When we perform the second derivative of our harmonic signal $h(t)$, we observe multiple morphologies. \autoref{fig:dominated_morphologies} depicts the three most dominant macro morphologies observed in $h''(t)$: $h_1''(t)$, $h_2''(t)$, $h_3''(t)$. 
Those dominant morphologies account for 98.4\% of the periods measured in a \textit{public dataset} with 35 subjects~\cite{PPG2019dataset}, 15301 out of 15557 periods
and 97.5\% of the periods measured in a \textit{private dataset} with 43 subjects, 11328 out of 11617 periods.
\autoref{fig:frequency_morphologies} shows the relative presence of those three macro morphologies in those datasets. There are other macro morphologies in $h''(t)$, but we do not consider them because they are rarely present (``Discard'' label).

If we would choose only one morphology as the template for all subjects, as conventional methods do, the system could face two major problems. \textit{First}, it may take a long time to identify a subject because the system will need to wait for the right morphology to arrive. For example, for the public dataset, our system can obtain 1.229 (15301/12444) periods/s,
compared to much lower speeds if would only use $h_1''(t)$ (0.3584 periods/s), $h_2''(t)$ (0.7427 periods/s), or  $h_3''(t)$ (0.1283 periods/s). \textit{Second}, and perhaps more critical, the right morphology may never arrive for one of the subject(s) or it may be so rare that there would be insufficient samples to train the system properly. 
In practice, such a limitation would render an identification system futile because the basic premise is that it should be able to identify \textit{all} members in a target group. In uncontrolled scenarios, no morphology is dominant. Even $h_2''(t)$, which is the most common, may be rarely active in some subjects, such as user 31 in the public dataset, and subjects 5 and 8 in the MIMIC-III dataset. Hence, the key advantages of considering multiple morphologies are: decreasing latency, and eliminating the risk of excluding some types of subjects.

\subsection{Feature extraction}

\begin{figure}[t!]
  \centering
    \begin{subfigure}{.48\columnwidth}
      \centering
      \vspace{-2mm}
      \includegraphics[width=.9\linewidth]{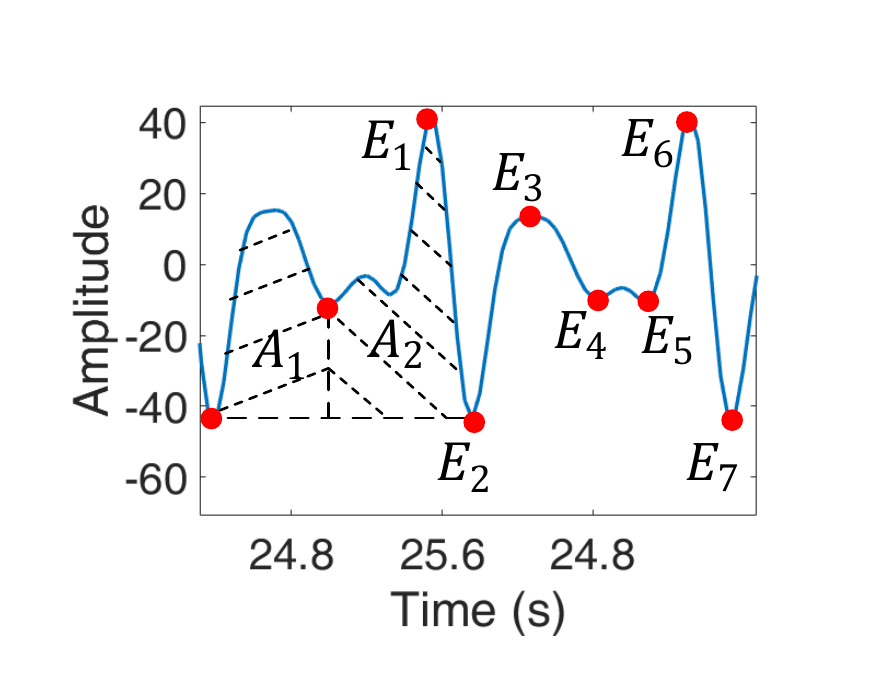} 
      \vspace{-2mm}
      \caption{Fiducial points on $h(t)$}
      \label{fig:HR_features}
    \end{subfigure}
    \hfill
    \begin{subfigure}{.48\columnwidth}
      \centering
      \vspace{-2mm}
      \includegraphics[width=.9\linewidth]{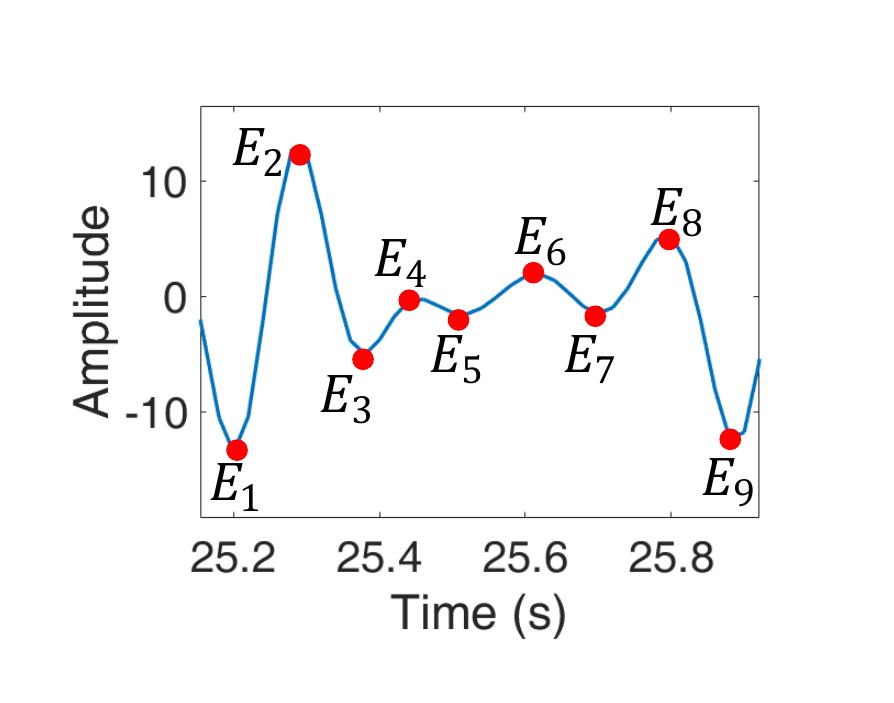} 
      \vspace{-2mm}
      \caption{Fiducial points on $h_2''(t)$}
      \label{fig:SD_features}
    \end{subfigure}
    \vspace{-2mm}
    \caption{Fiducial points used to extract features.}
    \vspace{-2mm}
    \label{fig:feature}
\end{figure}

We extract features from $h(t)$ and $h''(t)$. Like several other studies in the area~\cite{kavsaouglu2014novel,lin2017cardiac,liu2019cardiocam}, our features are largely based on the geometric relations amongst fiducial points. \autoref{fig:feature} and \autoref{tab:features} provide a pictorial representation and the notation for all the features. In our notation, $E_i(t)$ and $E_i(a)$ denote the time and amplitude of fiducial point $i$.
For $h(t)$, shown in \autoref{fig:HR_features}, we collect three types of features: \emph{1)} the duration of a period, \emph{2)} the ratio of the areas inside a period, and \emph{3)} the differences in duration, height and slope between consecutive fiducial points in one period. The total number of features for $h(t)$ is 14. For $h''(t)$, we only consider the third type of features (differences between contiguous fiducial points).  
\autoref{fig:SD_features} shows the features for $h''_2(t)$, and the same principle is applied to extract the features from $h''_1(t)$ and $h''_3(t)$. In the end, the number of features for $h_1''(t)$, $h_2''(t)$ and $h_3''(t)$, are 18, 24 and 30, respectively. 
Features based on duration and height are susceptible to heartbeat variance. In~\cite{israel2005ecg,liu2019cardiocam}, the authors state that normalizing the features makes them immune to heart rate changes. Therefore, we also normalized the duration and height features of $h(t)$ and $h''(t)$.

It is important to highlight that for $h(t)$ we also observed two types of morphologies: one where $E_4$ and $E_5$ are present and the other where those two points merge into a single valley. However, contrary to the multi-morphology approach used for $h''(t)$ in~\autoref{subsec:morphology_classification}, we decide to use a single morphology for $h(t)$ because the features obtained from the fiducial points $E_4$ and $E_5$ did not have any impact on the accuracy of the system. The information from those two points gets disentangled and captured in one of the three morphologies present in $h''(t)$. Due to this reason, we do not evaluate $i=4$ for $h(t)$ in \autoref{tab:features}.

\begin{table}[t]
\centering
\caption{Features extracted from $h(t)$ and $h''(t)$.}
\vspace{-2mm}
\renewcommand{\arraystretch}{1}
\scriptsize
\resizebox{\linewidth}{!}{
\begin{tabular}{|c|c|c|}
\hline
                                          & \textbf{Feature}                      & \textbf{Description}                                  \\ \hline
\multicolumn{1}{|c|}{\multirow{5}{*}{$h(t)$}} & \begin{tabular}[c]{@{}l@{}}Period duration\end{tabular}                           & $E_6(t)-E_1(t)$                   \\\cline{2-3} 
\multicolumn{1}{|c|}{}                    & Area ratio                    & $A_1/A_2$                                   \\ \cline{2-3} 
\multicolumn{1}{|c|}{}                    & \multicolumn{1}{c|}{Duration ($D_i$)} & $E_{i+1}(t)-E_{i}(t),i=2,3,5,6$ \\ \cline{2-3} 
\multicolumn{1}{|c|}{}                    & 
\multicolumn{1}{c|}{Height ($H_i$)}   & $|E_{i+1}(a)-E_{i}(a)|,i=2,3,5,6$ \\ \cline{2-3} 
\multicolumn{1}{|c|}{}                    & 
\multicolumn{1}{c|}{Slope}    & $(-1)^{i+1}*H_i/D_i$      \\ \hline

\multicolumn{1}{|c|}{\multirow{3}{*}{$h''(t)$}}                  & \multicolumn{1}{c|}{Duration ($D''_i$)} & $E''_{i+1}(t)-E''_{i}(t), 1 \leq i \leq 8$\\ \cline{2-3}
\multicolumn{1}{|c|}{}                    & 
\multicolumn{1}{c|}{Height ($H''_i$)}   & $|E''_{i+1}(a)-E''_{i}(a)|, 1 \leq i \leq 8$ \\ \cline{2-3}
\multicolumn{1}{|c|}{}                    & 
\multicolumn{1}{c|}{Slope}    & $(-1)^{i+1}*H''_i/D''_i$      \\ \hline
\end{tabular}
}
\vspace{-5mm}
\label{tab:features}
\end{table}

\section{Identification and Authentication}
\label{sec:iden_authen}

As stated earlier, SoA studies only evaluate one type of application: identification or authentication (mainly identification). We consider both. Our system relies on the same set of features for both cases. 
Upon receiving a raw PPG period, we first obtain $h(t)$ and $h''_i(t)$, and derive their features. The combined features, $h(t)$+$h''_i(t)$, are given as inputs to two different processing branches depending on the type of application. {Considering that performing identification is simpler, we first present that system, and later we focus on authentication.} 

\subsection{Identification}

Identification requires gathering training data from \textit{all subjects}, and during the testing phase the aim is to match an incoming cardiac sample to the right subject. As with many other classification problems, PPG identification requires two main components: \textit{dimensionality reduction}, to identify the most informative features; and \textit{decision boundaries}, to perform accurate classification.

\begin{figure}[t!]
    \centering
    \includegraphics[width=0.65\linewidth]{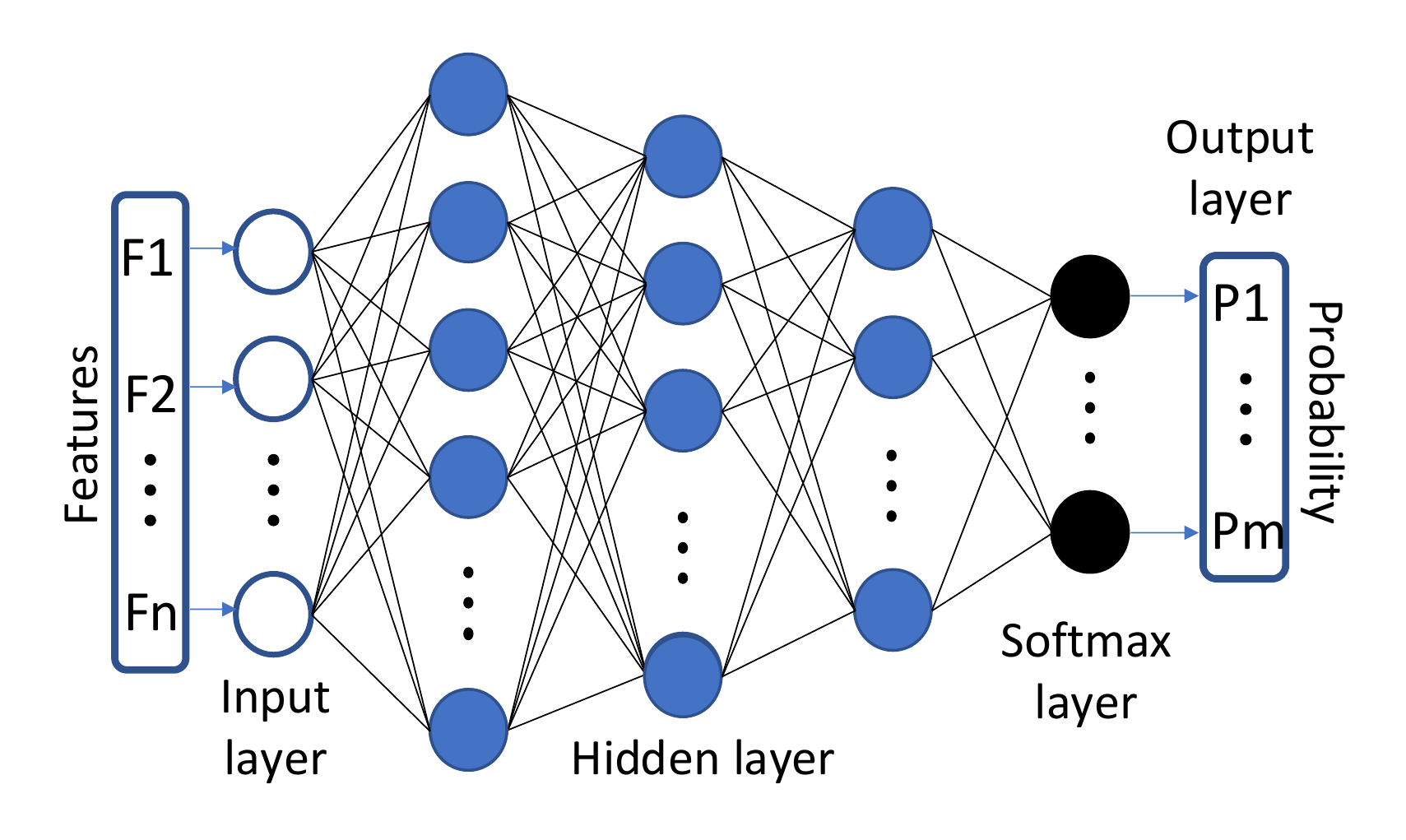}
    \vspace{-1mm}
    \caption{The structure of our neural network.}
    \label{fig:NN_structure}
\end{figure}

The SoA utilizes two kinds of supervised learning methods, linear and non-linear, but does not provide insights about which one is better and why. In our evaluation we consider both approaches. The most representative linear method is linear discriminant analysis (LDA) \cite{yadav2018emotion}, which simultaneously reduces dimensionality and draws decision boundaries. The most representative non-linear methods are based on neural networks (NN)~\cite{karimian2017human}. As it is customary with NN~\cite{hinton2006autoencoder}, we first use an autoencoder for dimensionality reduction, blue layers in~\autoref{fig:NN_structure}, and then, we add a softmax layer to perform classification (decision boundaries), {black} layer.

Considering that we use three morphologies, we need three LDA and NN pipelines running in parallel for each morphology (each pipeline receives the corresponding set of features presented in \autoref{tab:features}).
Since LDA is an analytical solution, the LDA module is the same for all three pipelines (but with different training data). 
In contrast to LDA, due to the influence of the network structure and parameter values, we tailor three different NN modules for each morphology.
The hidden (blue) layer neurons for morphologies one, two and three are 128-64-32, 170-85-42 and 128-64-32, respectively. The activation functions of neurons are sigmoid to guarantee the non-linearity of the system, and parameters such as L2 and sparsity regularization are tuned for each morphology.

\subsection{Authentication}
Contrary to identification, in authentication systems, the training set only consists of samples from the \textit{legitimate subject}, while its testing set can include samples from \textit{any subject}. Authentication also requires dimensionality reduction 
(to identify informative features and reduce the number of features in case of curse of dimensionality) 
and boundaries, but given that we lack information about other users, drawing an optimal boundary for that single legitimate user becomes more complex. Next, we first explain the methods used in the SoA for dimensionality reduction, and then, some techniques to improve the definition of boundaries.

\textbf{Dimensionality reduction.}
It can also be performed with linear and non-linear methods. There are two mainstream linear techniques: principal component analysis (PCA)~\cite{pearson1901pca} and non-negative matrix factorization (NMF)~\cite{lee1999NMF}. NMF requires non-negative features, but the slopes in our feature set can be negative. Hence, similar to prior studies~\cite{liu2019cardiocam}, we also adopt PCA. 
Even though there are several non-linear dimensionality reduction techniques --such as Isomap \cite{tenenbaum2000isomap}, local linear embedding (LLE) \cite{roweis2000LLE}, t-distributed stochastic neighbor embedding (t-SNE) \cite{maaten2008tSNE}, and autoencoder \cite{hinton2006autoencoder}-- we did not find SoA studies using them for PPG authentication. Isomap, LLE and t-SNE share a common disadvantage for PPG-authentication: they must perform an entire recalculation every time a new test point is added. Autoencoders, on the other hand, do not have that shortcoming.  We performed a preliminary evaluation of authentication with autoencoders but the performance was not good. We hypothesize that it is due to the limited data, autoencoders are usually trained with at least thousands of training points\footnote{At a popular Quora forum discussing ``How large should be the data set for training a Deep auto encoder?" Yoshua Bengio states the need for having large amounts of training data \cite{YoshuaBengio}.}. PPG-based systems are trained with a few minutes of cardiac data in one subject, which maps to a few hundred cardiac periods. In identification, NN counts with several thousand samples coming from \textit{all users}, but in authentication, we only have a few hundred samples coming from \textit{the legitimate user}. Due to this finding, in our evaluation section, we only consider PCA for authentication. 

\begin{figure}[t]
    \vspace{-3mm}
    \centering
    \begin{minipage}{.45\columnwidth}
      \includegraphics[width=0.9\linewidth]{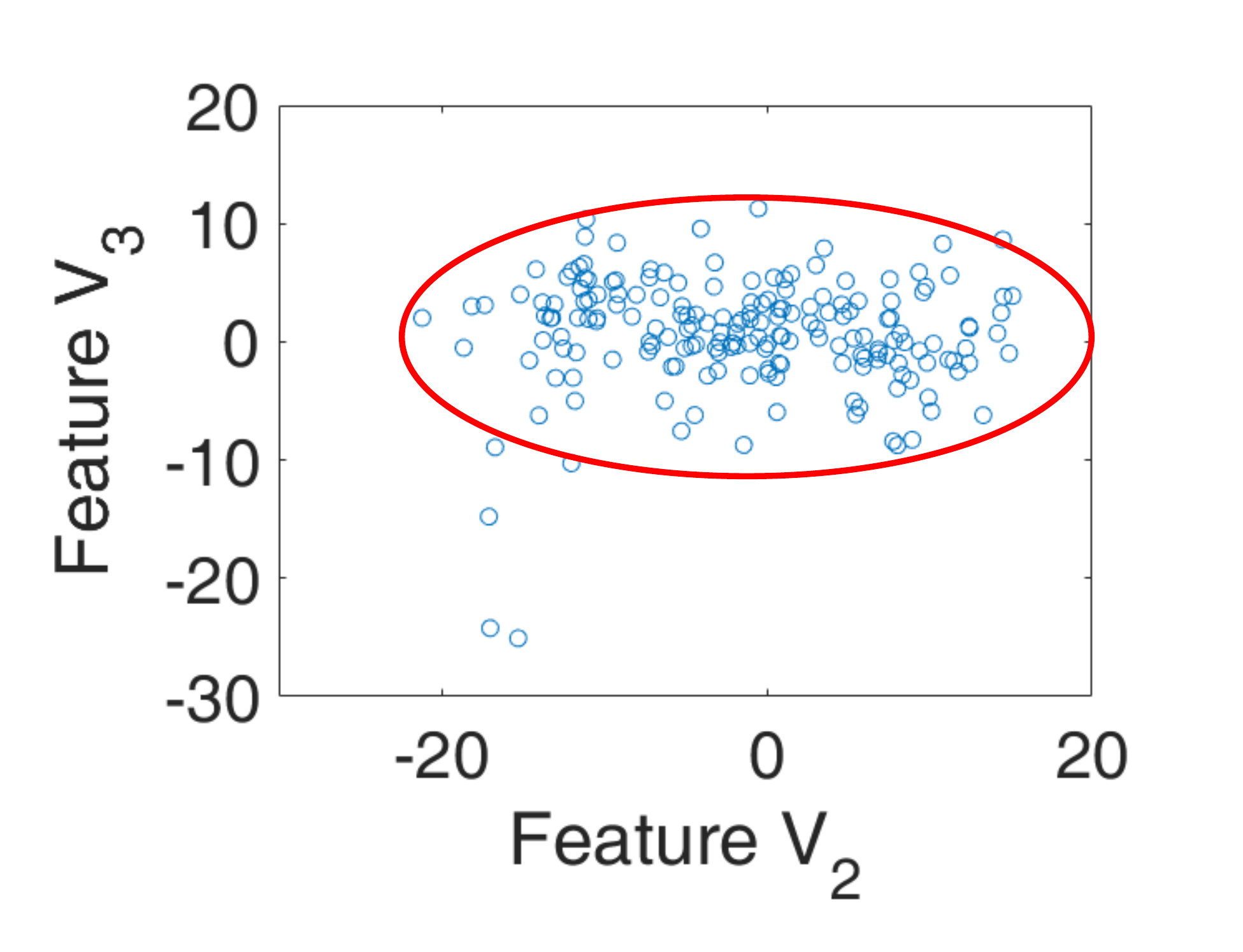}
      \vspace{-2mm}
      \caption{Mahalanobis.}
      \vspace{-1mm}
      \label{fig:mahal_reason}
    \end{minipage}
    \hfill
    \begin{minipage}{.45\columnwidth}
        \centering
        \includegraphics[width=0.9\linewidth]{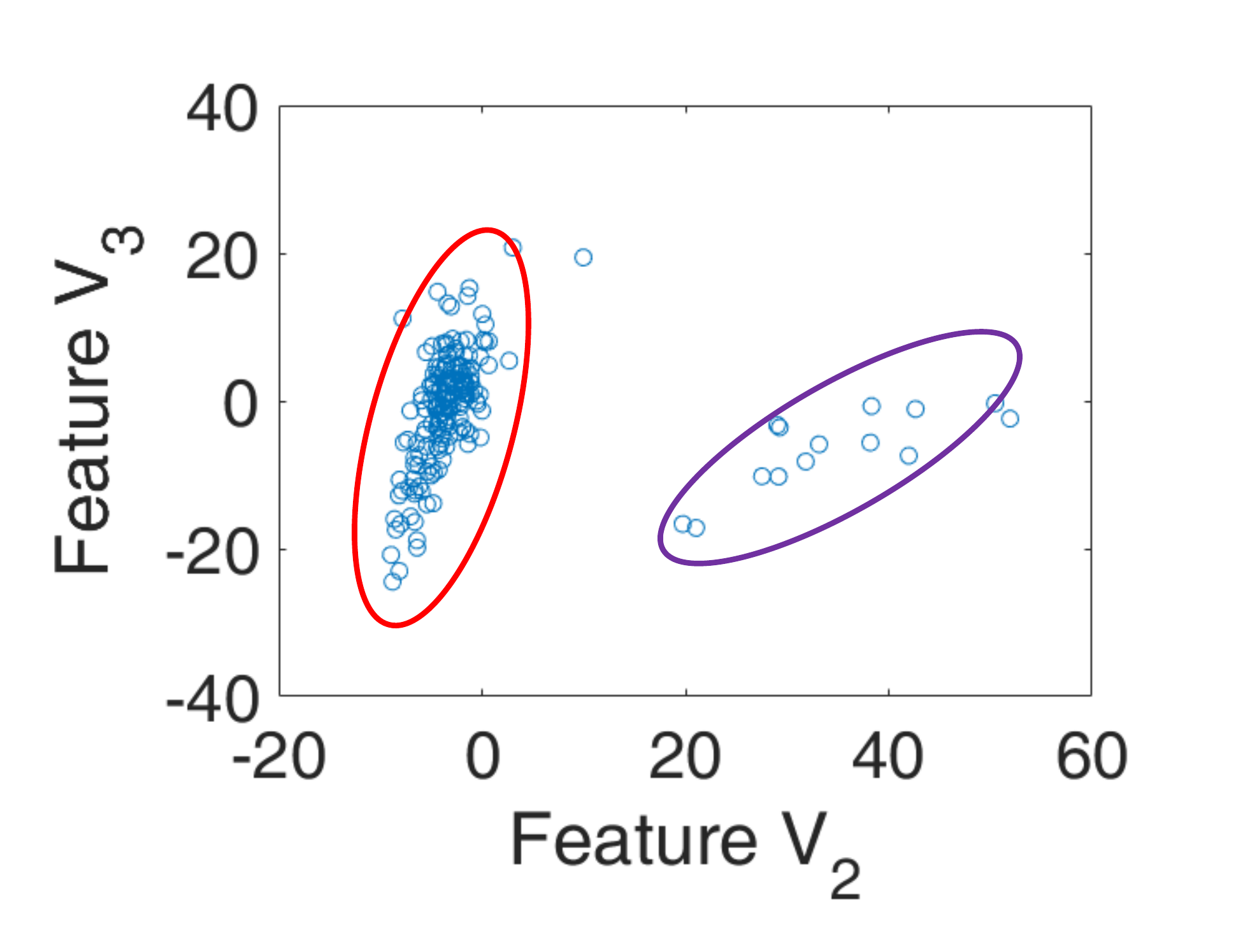}
        \vspace{-2mm}
        \caption{Multi-cluster.}
        \vspace{-1mm}
        \label{fig:cluster_inter_reason}
    \end{minipage}
\end{figure}

\textbf{Mahalanobis distance.}
After dimensionality reduction, the most significant features of a subject usually form a cluster similar to the one shown in~\autoref{fig:mahal_reason}. When a new test sample arrives, the system calculates the average distance of this new point to the cluster. If the distance is below a threshold, the user is deemed legitimate. 
Many studies use Euclidean distances to measure proximity~\cite{liu2019cardiocam}. But Euclidean distances are fundamentally ill-equipped to deal with feature spaces that have widely different variances. 
For example, in~\autoref{fig:mahal_reason}, using Euclidean distances, with any threshold, leads to a boundary that has the shape of a circle. The circle will be either too long for $v_3$, causing numerous false positives; or too short for $v_2$, causing significant false negatives. Therefore we adopt the well-known Mahalanobis distance~\cite{mahalanobis1936generalized}, which considers the standard deviations in each dimension and can be used to define tight boundaries such as the red ellipsoid shown in~\autoref{fig:mahal_reason}.

\textbf{Multi-cluster approach.}
Current PPG authentication systems assume that the features of a user converge to a \textit{single} cluster~\cite{yadav2018emotion,liu2019cardiocam}. 
However, with uncontrolled data, we observed that a single subject can form two or more clusters for a single morphology, as depicted in~\autoref{fig:cluster_inter_reason}. We need an authentication system that can identify multiple clusters and then use the Mahalanobis distance to set an appropriate threshold for each cluster.\footnote{To calculate the Mahalanobis distance, the number of samples must be greater than the number of features (dimensions). If a cluster has few samples, like the purple one in~\protect{\autoref{fig:cluster_inter_reason}}, we use spline interpolation to add the extra necessary points.}

For our purposes, the clustering method should meet three requirements: 
\emph{(i)} be resilient to the presence of outliers, \emph{(ii)} able to detect clusters with arbitrary shape, and \emph{(iii)} fast.
Hierarchical clustering methods, such as BIRCH~\cite{peng2018BIRCH}; and centroid-based methods like K-means~\cite{macqueen1967kmean} are vulnerable to outliers and cannot detect arbitrary shapes. 
Most grid-based clustering methods, like CLIQUE~\cite{agrawal1998CLIQUE}, and density-based methods, like OPTICS~\cite{ankerst1999optics} and DBSCAN~\cite{ester1996DBSCAN} do not have shortcomings (i) and (ii), but they need a relatively long computation time. 
Due to the above reasons, we decide to use WaveCluster~\cite{sheik2000wavecluster}, which exploits the multi-resolution property of wavelet transforms. WaveCluster
can identify arbitrary shape clusters at different degrees of accuracy, it is insensitive to outliers and has a low time complexity $O(n)$.

\section{Performance Evaluation}
\label{sec:eval}

In this section, we describe the datasets we use, the studies taken from the SoA as baselines for comparison, and the results for the evaluation of identification and authentication.


\subsection{Datasets}
\label{subsec:datasets}

We use two datasets to evaluate the performance of CardioID.
The first dataset, using \textit{pulse oximeter} for signal collection, is public and recently published in~\cite{PPG2019dataset}. All subjects are sitting during the signal collection.
The second dataset is collected by us.\footnote{This data collection is performed under the approval from the HERC of our university.}
We use an iPhone 7 \textit{camera} to gather cardiac periods from sitting volunteers.
The camera records videos of the redness change on our volunteers' fingertips attached to the camera at 60 FPS. In each frame, we focus only on the red channel of the pixels covered by the fingertip. The method to select the covered pixels is the same as in~\protect{\cite{liu2019cardiocam}}: $I_{\text{red}}  > 80\% \times (I_{\text{red}}+I_{\text{blue}}+I_{\text{green}})$. Then, we average the red-channel intensities among the selected pixels to represent one data point of a PPG signal.
To maximize the peak-to-peak amplitude of cardiac periods, we carefully set the three parameters affecting the camera's exposure: the aperture and ISO are set to the lowest values, -2 and 20, respectively, and the shutter speed to 200. Other parameters like white balance, focus and zoom are set as auto. 


\begin{table}[t]
\centering
\caption{Details of the three datasets used in our evaluation.}
\vspace{-2mm}
\label{tab:dataset_situation}
\resizebox{\linewidth}{!}{
\renewcommand{\arraystretch}{1}
\begin{tabular}{|c|c|c|c|c|c|}
\hline
                                                                  & \textbf{\# subjects} & \textbf{\# female} & \textbf{\begin{tabular}[c]{@{}c@{}}Age (Mean,\\ Variance, Range)\end{tabular}} & \textbf{\begin{tabular}[c]{@{}c@{}}RD\\ (mins)\end{tabular}} & \textbf{CTE}\\ \hline
\begin{tabular}[c]{@{}c@{}}Public (pulse\\ oximeter)\end{tabular} & 35                   & 12                 & \begin{tabular}[c]{@{}c@{}}28.4, 14.04,\\ 10-75 Y/O\end{tabular}               & 5-6      & 4.29          \\ \hline
\begin{tabular}[c]{@{}c@{}}Private \\ (camera)\end{tabular}       & 43                   & 16                 & \begin{tabular}[c]{@{}c@{}}36.7, 14.93,\\ 12-79 Y/O\end{tabular}               & 4      & 5.97            \\ \hline
\end{tabular}
}
\vspace{-6mm}
\end{table}

\vspace{1mm}
\textbf{Significance of datasets.} The parameters of the three datasets --the number of subjects, gender, age distribution (average, standard deviation and range), the recording duration (RD) and variability (cross-track error, CTE)-- are given in \autoref{tab:dataset_situation}. 
There are two important points to highlight about the selection of our datasets. 

First, \textit{no SoA study has analyzed the performance of their methods using both types of sensors, pulse oximeter and camera}. In general, a pulse oximeter is more precise than a camera because its infrared spectrum can enhance the signal quality, and its finger clip can reduce the noisy motion artifacts~\cite{Farooq2010segment}. This is one reason why the CTE (variability) in \autoref{tab:dataset_situation} is higher for the camera dataset.

Second, \textit{our first two datasets consider a broader group of people, which allows us to consider a wide range of dynamics and prevent the system from overfitting a narrow segment of users.}  
Even for \textit{healthy} people, which is the main focus of our work and most of the SoA, the morphology of cardiac signals can vary significantly based on the age group and skin tone. This is another reason why the CTE in \autoref{tab:dataset_situation} is higher for the camera dataset.
Our baseline studies, \cite{liu2019cardiocam} and \cite{kavsaouglu2014novel}, focus on a narrow age segment of the adult population, 22-33 and 18-46, respectively. The results obtained with these concentrated age distributions are hard to validate for people of all ages.
The age ranges of our datasets are 10-75 (public) and 12-79 (private), including children, teenagers, adults and elders. Moreover, we also consider a bigger population: 70\% more for the camera sensor (43 vs. 25, \cite{liu2019cardiocam}) and 17\% more for the pulse oximeter (35 vs. 30, \cite{kavsaouglu2014novel}). In terms of skin tones, the public dataset includes only medium-toned skin, and our private dataset includes 51\% medium, 33\% light and 16\% dark skin.


\begin{figure}[t]
    \centering 
    \vspace{-3mm}
        \begin{subfigure}{.47\columnwidth}
          \centering
          \includegraphics[width=\linewidth]{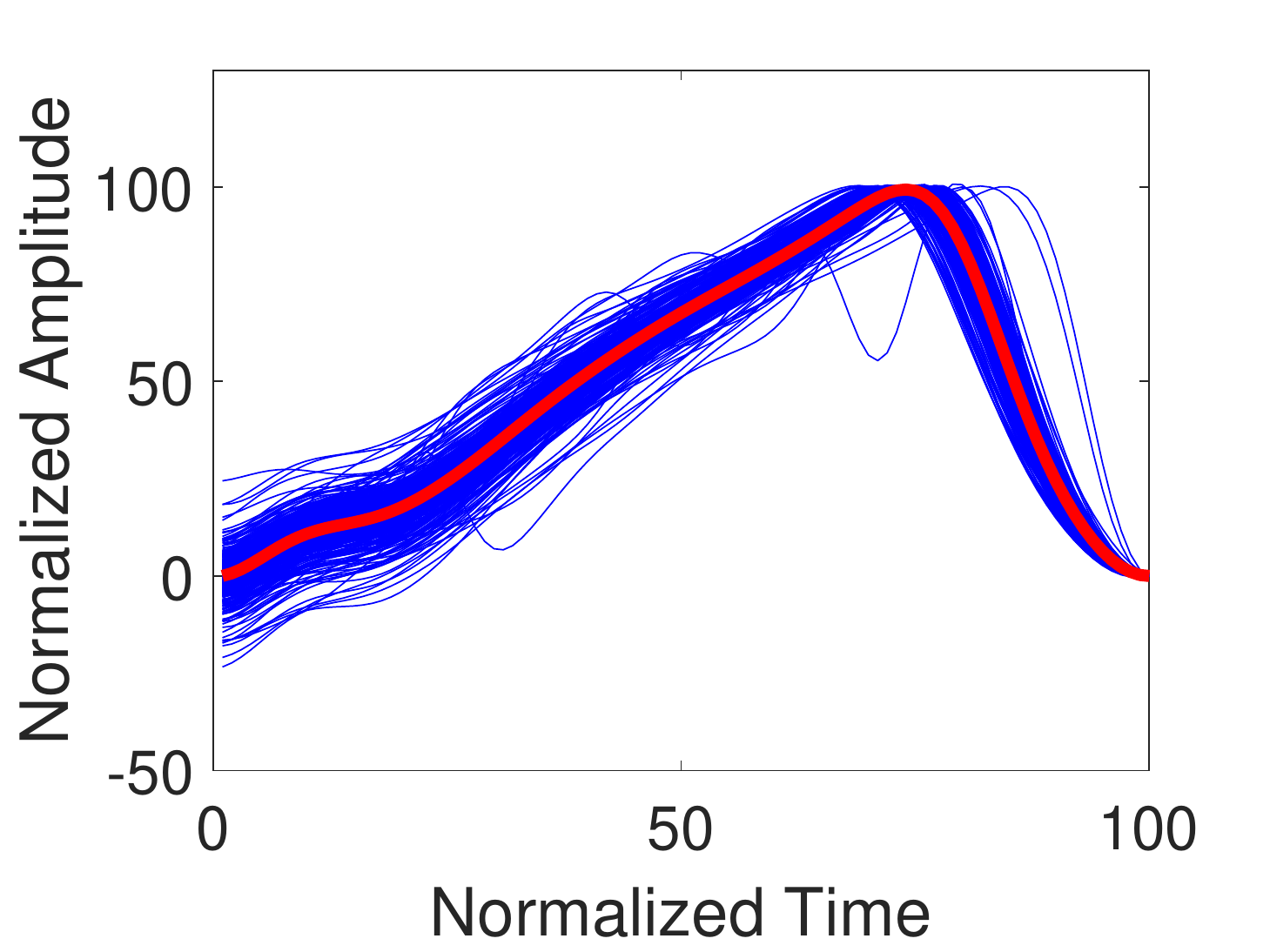}
          \vspace{-5mm}
          \caption{Variance 1.8}
          \vspace{-1mm}
        \label{fig:variance_2}
        \end{subfigure}\hfil 
        \begin{subfigure}{.47\columnwidth}
          \centering
          \includegraphics[width=\linewidth]{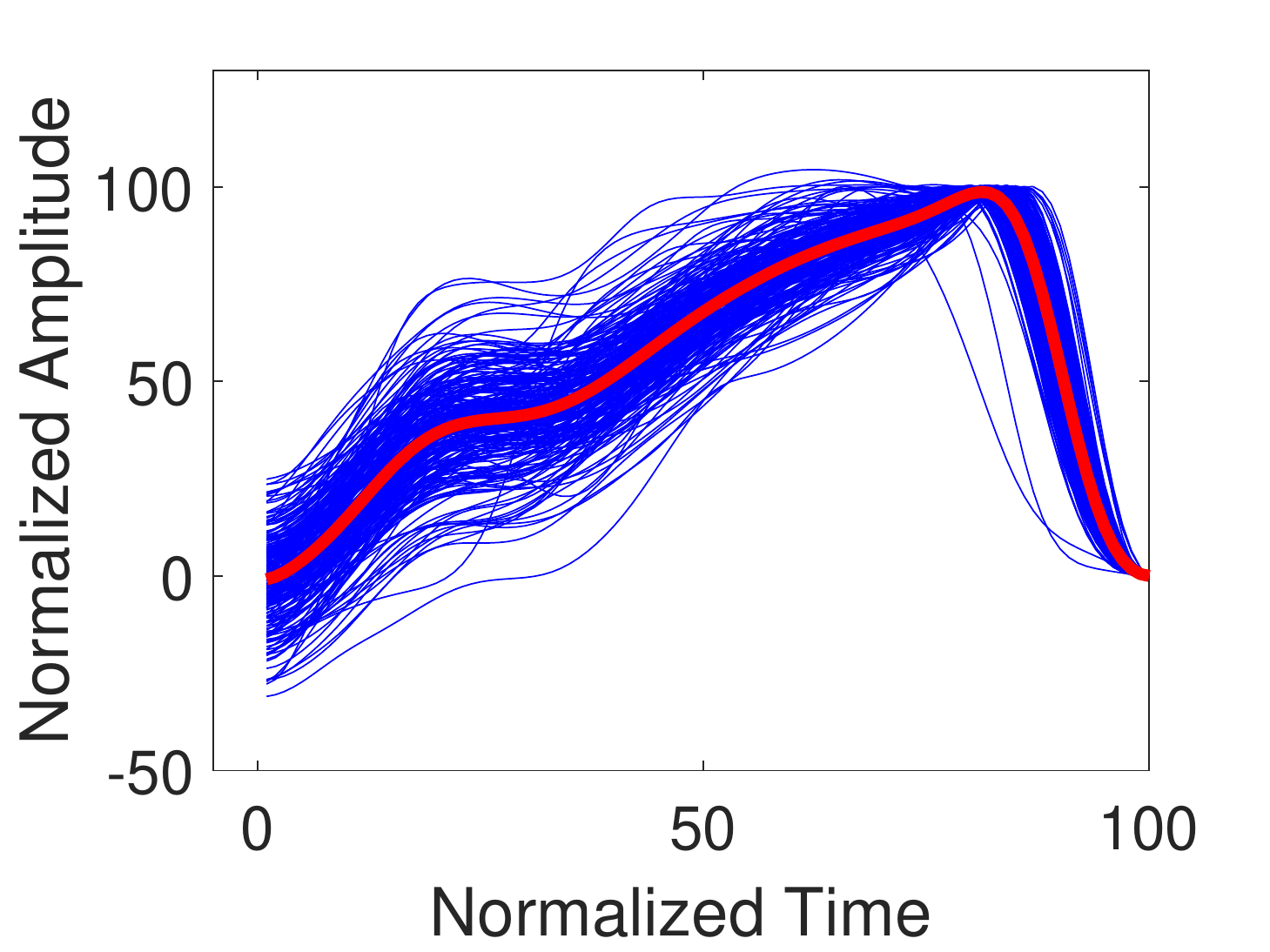}
          \vspace{-5mm}
          \caption{Variance 3.9}
          \vspace{-1mm}
        \label{fig:variance_4}
        \end{subfigure}
        
        \medskip
        \vspace{-2mm}
        \begin{subfigure}{.47\columnwidth}
        \centering
        \includegraphics[width=\linewidth]{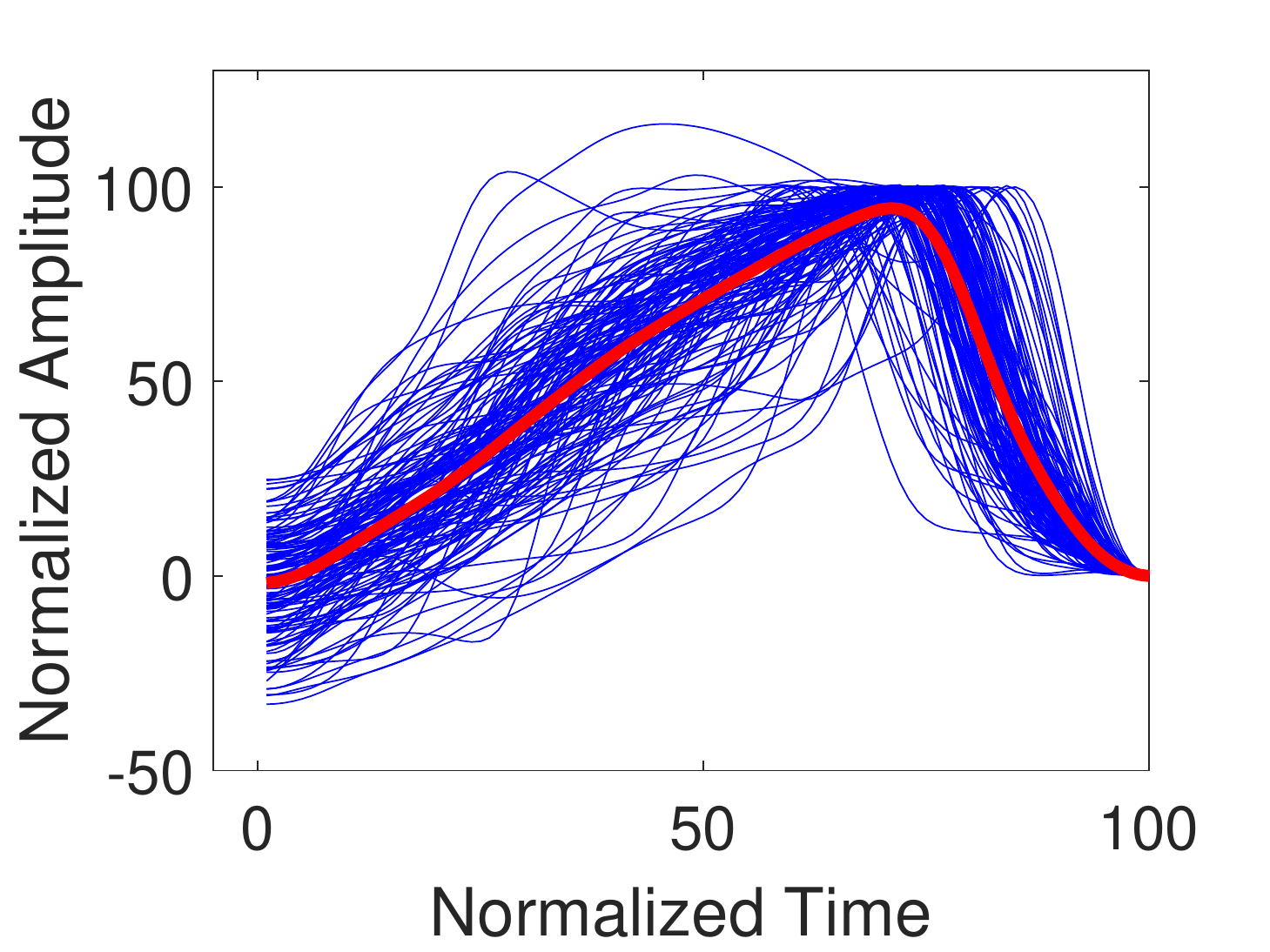}
        \vspace{-5mm}
        \caption{Variance 5.8}
        \label{fig:variance_6}
        \end{subfigure}\hfil 
        \vspace{-2mm}
        \begin{subfigure}{.47\columnwidth}
        \centering
        \includegraphics[width=\linewidth]{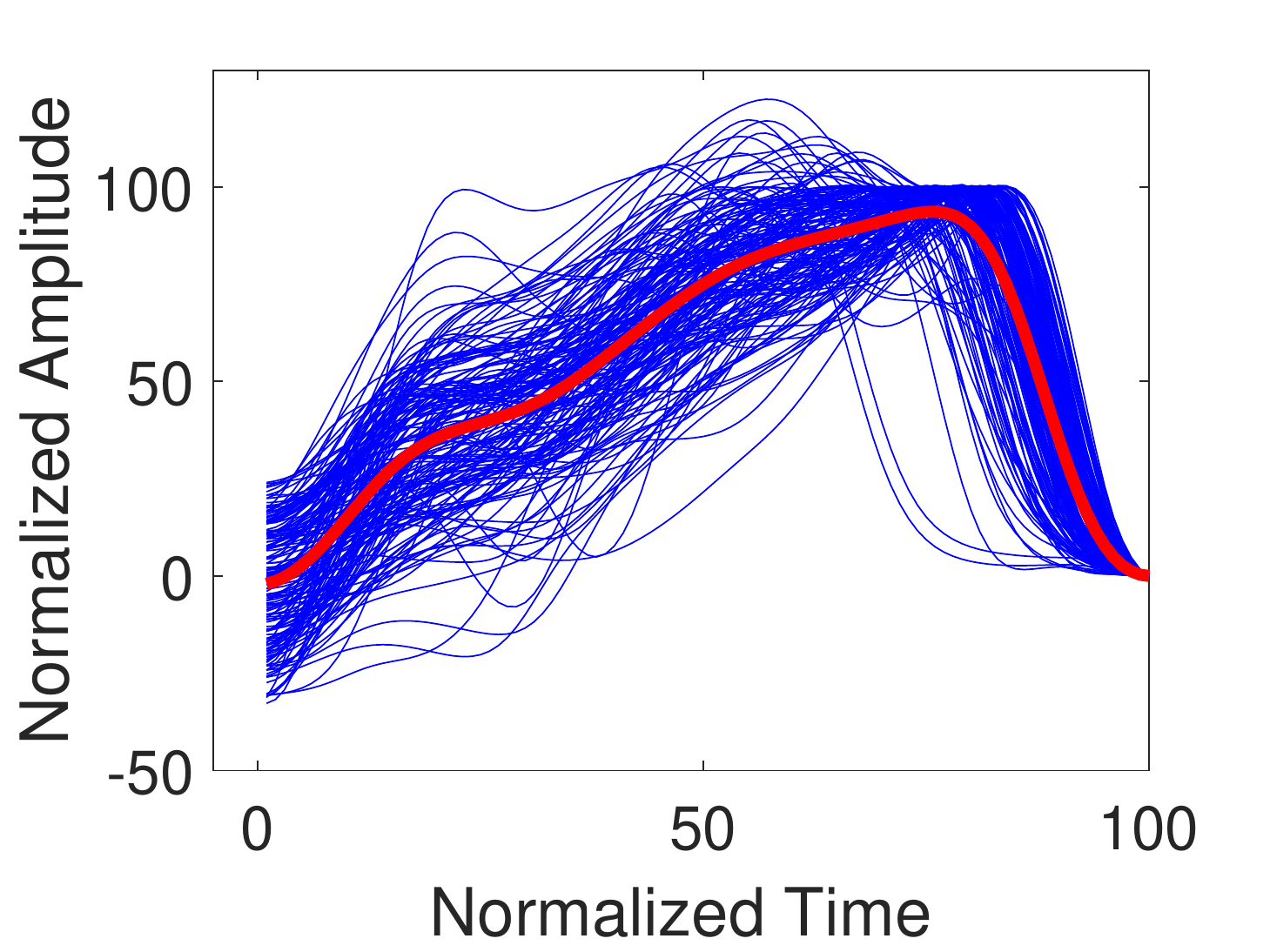}
        \vspace{-5mm}
        \caption{Variance 6.6}
        \label{fig:variance_6_plus}
        \end{subfigure}
    \caption{Different signal variances.}
    \label{fig:signal_variance}
    \vspace{-5.5mm}
\end{figure}

\textbf{Signal variance analysis.} \label{subsubsec:CTE} Gathering data from different sensors, while considering motion artifacts and a broader range of people, provides us with more realistic (less controlled) PPG signals. To quantify the variance of these signals, we first obtain the average signal for a user, red signals in~\autoref{fig:signal_variance}; and then, we calculate the cross-track error (CTE)\footnote{The CTE is used in GPS systems to measure the difference between the given and followed paths. We tried different similarity metrics, including dynamic time warping, and we found that the CTE captures the similarity of PPG signals in a more precise manner.} from every (blue) PPG signal to its average. Denoting $e_i$ as the CTE for signal $i$, the signal variance for a subject is the mean absolute error for all $e_i$'s.
The average signal variances for our datasets are 4.29 and 5.97,
shown in~\autoref{tab:dataset_situation}. \textit{In order to put these values in context, it is important to note that the variance found in SoA plots is a bit lower than what is shown in~\protect{\autoref{fig:variance_2}}, less than 2}. The majority of users in the public dataset have a variance in the range $[2,6]$ and in private dataset in $[4,6]$ 
Hence, our evaluation copes with a spectrum of signal variability that has not been tackled before.

\subsection{Baselines used for comparison}
\label{subsec:baselines}

We utilize two SoA studies as baselines for comparison, one for identification~\cite{kavsaouglu2014novel} and the other for authentication~\cite{liu2019cardiocam}. The reasons for selecting those baselines are presented in \autoref{sec:soa}. In this subsection, we quantity the acquisition rate improvement of \protect{\cardioid} and the performance of our work and the SoA baselines.

\subsubsection{Quantifying acquisition rates} Every study in this research area, including ours, removes periods that do not conform to the required morphologies. The goal is to discard as few periods as possible, while maintaining high accuracy. Denoting $S$ as the cardiac periods from all users and $S'$ as the periods used by a system (after discarding non-conforming morphologies), the acquisition rate is given by $S'/S$. With controlled data, the acquisition rate is high, $S' \approx S$; but with uncontrolled data, the rate can be very low, $S' \ll S$. As stated in~\autoref{sec:feature}, a low rate can increase the system's delay and in some cases exclude the participation of some users. Hence, before even assessing the accuracy of the system, we need to make sure that a method has the capability to recognize 100\% of the users.

Considering that $S$ are equal to 14347 and 10728
periods for the public and private
datasets, respectively, we first need to find $S'$ for the SoA baselines. The morphology and features used by Kavsaouglu \textit{et. al.} are presented in Figures 7 and 8 in~\cite{kavsaouglu2014novel}, and the corresponding information for CardioCam is provided in Figure 8 in~\cite{liu2019cardiocam}.\footnote{Upon close inspection, we notice that, in both SoA studies, the second derivative of $f(t)$ is a signal similar, but not exactly the same, as morphology-2 in our case, $h''_2(t)$.} 
We use that information to discard the morphologies that do not conform to their requirements (the right morphology is necessary to obtain their features). After discarding the non-conforming morphologies in SoA, we obtain the following acquisition rates: 74.6\% (public dataset) and 64.5\% (private dataset)
for \cite{kavsaouglu2014novel}; and 59.2\% (public) and 32.8\% (private)
for \cite{liu2019cardiocam}; significantly lower than the 98.4\% (pulse oximeter) and 97.5\% (camera)
obtained for \cardioid. \textit{More importantly, in the camera dataset there were three users that \textbf{did not have a single cardiac period} resembling the morphology required by \cite{liu2019cardiocam}, and thus, there is no possibility to authenticate them with that method.}

Moreover, the solutions in these SoA \cite{kavsaouglu2014novel,liu2019cardiocam} have a long acquisition delay. There are 12444, 10320, and 3000 seconds for the public dataset, private dataset and MIMIC-III dataset, respectively. The acquisition speeds are 0.932 (public dataset) and 0.726 (private dataset)
periods/second for \cite{kavsaouglu2014novel}; and 0.740 (public) and 0.369 (private)
periods/second for \cite{liu2019cardiocam}; significantly lower than the 1.229 (pulse oximeter) and 1.097 (camera)
periods/second obtained for \cardioid. Those SoA acquisition speeds are below 1 period/second. Some of the acquisition speeds are even lower than 0.5 period/second, which is unfriendly for users.

\subsubsection{\cardioid variants} To assess the impact of our contributions --\textit{morphology stabilization}, \textit{morphology classification} and \textit{the reduction of non-linear effects for authentication}-- we create different variants for \cardioid. SoA approaches are implemented based on morphologies and features provided in their respective studies.

For identification, we consider four variants.
\begin{itemize}
    \item Variant I.1 (MS): we use \textbf{morphology stabilization} to obtain $h(t)$ and $h''(t)$ with their respective features, cf. \autoref{tab:features}. This variant only considers morphology-2 periods, $h''_2(t)$. The classification method is K-NN, the same as in \cite{kavsaouglu2014novel}.
    \item Variant I.2 (MC): we add \textbf{morphology classification} to the MS variant. Here, we include periods with morphologies $h''_1(t)$ and $h''_3(t)$ but we still use K-NN for the classification.
    \item Variants I.3 and I.4 (\cardioid.LDA and \cardioid.NN): we replace K-NN in the MC variant with \textbf{LDA} and \textbf{NN}, respectively. We consider these variants the final implementations of \cardioid for identification.
\end{itemize}

For authentication, we also consider four variants. 
\begin{itemize}
    \item Variant A.1 (MS): similar to the MS variant used for identification, but instead of K-NN, it uses PCA and Euclidean distance to achieve authentication, as in \cite{liu2019cardiocam}.
    \item Variant A.2 (MC): it adds \textbf{morphology classification} to the MS variant.
    \item Variant A.3 (Mahal): it replaces the Euclidean distance with \textbf{Mahalanobis distance} in the MC variant.
    \item Variant A.4 (\cardioid): it adds the \textbf{multi-cluster} approach to the Mahal variant.
\end{itemize}

\subsubsection{Emulating a wide range of signal variances} Our aim is to evaluate the SoA baselines and \cardioid variants under a wide range of signal variances. Collecting that type of data would require asking users to steadily increase the level of finger movement and pressure from low to high. That would be a complex process, instead we decide to divide our datasets to create (emulate) subsets with increasing levels of variance. To generate the emulated subsets, we perform the following process. For every user, we only include signals that lead to a variance less than $t$, where $t=2,4,6$. If after this filtering process, a variant cannot collect 20 periods from a user, we leave the user out of the emulated set because we would not have sufficient training data for that user. A macro view of the emulated subsets is presented in \autoref{Tab:online_periods} and \autoref{Tab:camera_periods}.
For example, if we look at \autoref{Tab:online_periods} for reference \cite{kavsaouglu2014novel}, we can see that if we set $t=2$, \emph{(i)} only 21.6\% of the periods would have a variance less than 2 \textbf{and} satisfy the morphology requirement of \cite{kavsaouglu2014novel}; and \emph{(ii)} only 29 subjects, out of 35 (82.9\%), have more than 20 periods satisfying the conditions in (i).   

\autoref{Tab:online_periods} and \autoref{Tab:camera_periods} provide two important insights. First, for all variance levels, the MC variant has the best performance in terms of including more users and having the highest acquisition rate (because it accepts three different morphologies). The SoA baselines and the MS variant have lower performance because they consider only one morphology. 
Second, when we consider all the data (last column in the tables), one of the baselines~\cite{liu2019cardiocam} cannot include all users in both datasets. 
This is an important point showing that the requirements in \cite{liu2019cardiocam} for the single morphology is so stringent that some users may rarely (or never) show the required morphology.
\textit{In fact, with the camera dataset, three users did not have a single cardiac period satisfying the morphology requirements, and two of those users were above 50 years old (an age bracket that was not considered in SoA)}. The broad type of users in our dataset enables us to expose this age limitation. 
The variants, LDA, NN, Mahal and Cluster, have the same \% of subjects and periods as MC because they are derived from that variant. 

\begin{table}[t]
\centering
\caption{Percentage of detectable subjects and periods in the public (pulse-oximeter) dataset.}
\vspace{-2mm}
\label{Tab:online_periods}
\renewcommand{\arraystretch}{1}
\resizebox{.9\columnwidth}{!}{
\begin{tabular}{|c|c|c|c|c|c|}
\hline
\textbf{Public}          & \textbf{Signal variance} & \textbf{2} & \textbf{4} & \textbf{6} & \textbf{All} \\ \hline
\multirow{2}{*}{\cite{kavsaouglu2014novel}} & Subject \%               & 82.9       & 100        & 100        & 100          \\ \cline{2-6} 
                         & Period \%                & 21.6       & 54.2       & 67.3       & 74.6         \\ \hline
\multirow{2}{*}{\cite{liu2019cardiocam}} & Subject \%               & 68.6       & 80         & 82.9       & 82.9         \\ \cline{2-6} 
                         & Period \%                & 19.5       & 45.9       & 55.2       & 59.2         \\ \hline
\multirow{2}{*}{MS}      & Subject \%               & 57.1       & 97.1       & 97.1       & 97.1         \\ \cline{2-6} 
                         & Period \%                & 8.1        & 29.8       & 44.9       & 59.4         \\ \hline
\multirow{2}{*}{MC}      & Subject \%               & 74.3       & 100        & 100        & 100          \\ \cline{2-6} 
                         & Period \%                & 14.8       & 50.6       & 75.1       & 98.39        \\ \hline
\end{tabular}}
\vspace{-2mm}
\end{table}

\begin{table}[t]
\centering
\caption{Percentage of detectable subjects and periods in the public (camera) dataset.}
\label{Tab:camera_periods}
\vspace{-2mm}
\renewcommand{\arraystretch}{1}
\resizebox{0.9\columnwidth}{!}{
\begin{tabular}{|c|c|c|c|c|c|}
\hline
\textbf{Private}         & \textbf{Signal variance} & \textbf{2} & \textbf{4} & \textbf{6} & \textbf{All} \\ \hline
\multirow{2}{*}{\cite{kavsaouglu2014novel}} & Subject \%               & 37.2       & 90.7       & 100        & 100          \\ \cline{2-6} 
                         & Period \%                & 5.5        & 33.1       & 51.2       & 64.5         \\ \hline
\multirow{2}{*}{\cite{liu2019cardiocam}} & Subject \%               & 25.6       & 41.9       & 58.1       & 69.8         \\ \cline{2-6} 
                         & Period \%                & 4.7        & 18         & 25.9       & 32.8         \\ \hline
\multirow{2}{*}{MS}      & Subject \%               & 18.6       & 81.4       & 95.3       & 100          \\ \cline{2-6} 
                         & Period \%                & 3.4        & 22.3       & 37.5       & 55.5         \\ \hline
\multirow{2}{*}{MC}      & Subject \%               & 30.2       & 93         & 100        & 100          \\ \cline{2-6} 
                         & Period \%                & 5.5        & 37.6       & 64.3       & 97.5         \\ \hline
\end{tabular}}
\vspace{-4mm}
\end{table}



\subsubsection{Providing the right context for evaluation } Until now, we have only evaluated the impact of \cardioid on the acquisition rate. In the remainder of this section we evaluate its performance, and it is important to consider the following points: (i) for each emulated subset, we use the first 80\% of periods for training and the rest for testing; (ii) the variance observed in SoA signals is around 1.5, hence, the variance considered in our emulated subsets ($t=2,4,6$) poses a greater challenge; (iii) for \cardioid and the SoA baselines, we use a \textit{single period} to perform identification or authentication, using more periods would increase the performance but also latency; (iv) using a single cardiac period with \textit{controlled signals}, the SoA reports a BAC of 0.95~\cite{liu2019cardiocam}, which can be translated, in expectation, to the rightful user having a probability around 95\% to get access to the system (sensitivity), and an attacker a probability of around 5\% of being successful (specificity). Thus, our goal is to try to approach 0.95 BAC with \textit{uncontrolled signals}. In authentication and identification, \textit{improvements in the order of 5\%, or above, are already considered highly significant}.

\subsection{Identification} 

\begin{figure}[t!]
    \begin{subfigure}{.495\columnwidth}
      \centering
      \includegraphics[width=\linewidth]{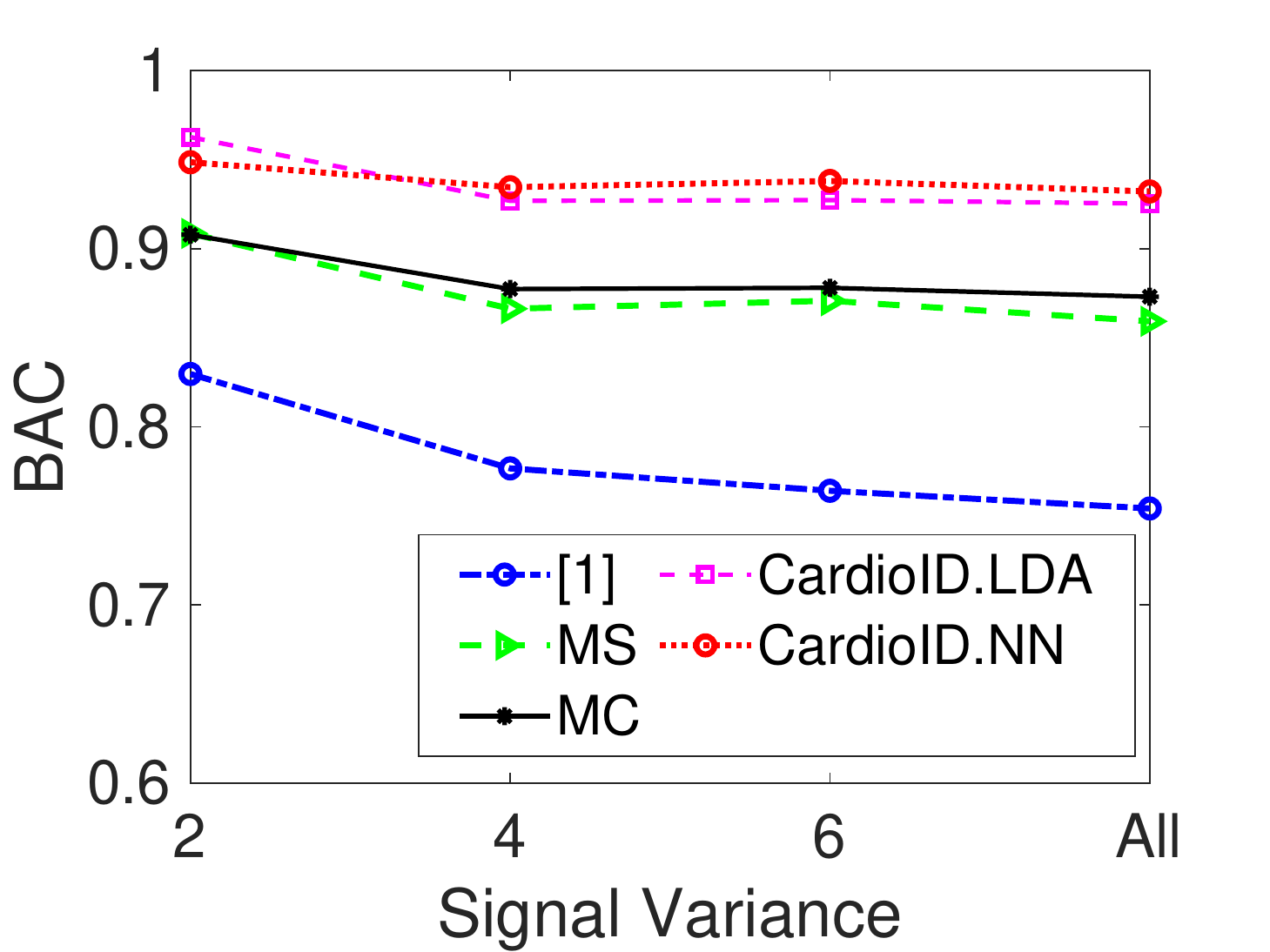}
      \vspace{-5mm}
      \caption{Pulse oximeter dataset}
    \label{fig:online_iden}
    \end{subfigure}
    \hfill
    \begin{subfigure}{.495\columnwidth}
      \centering
      \includegraphics[width=\linewidth]{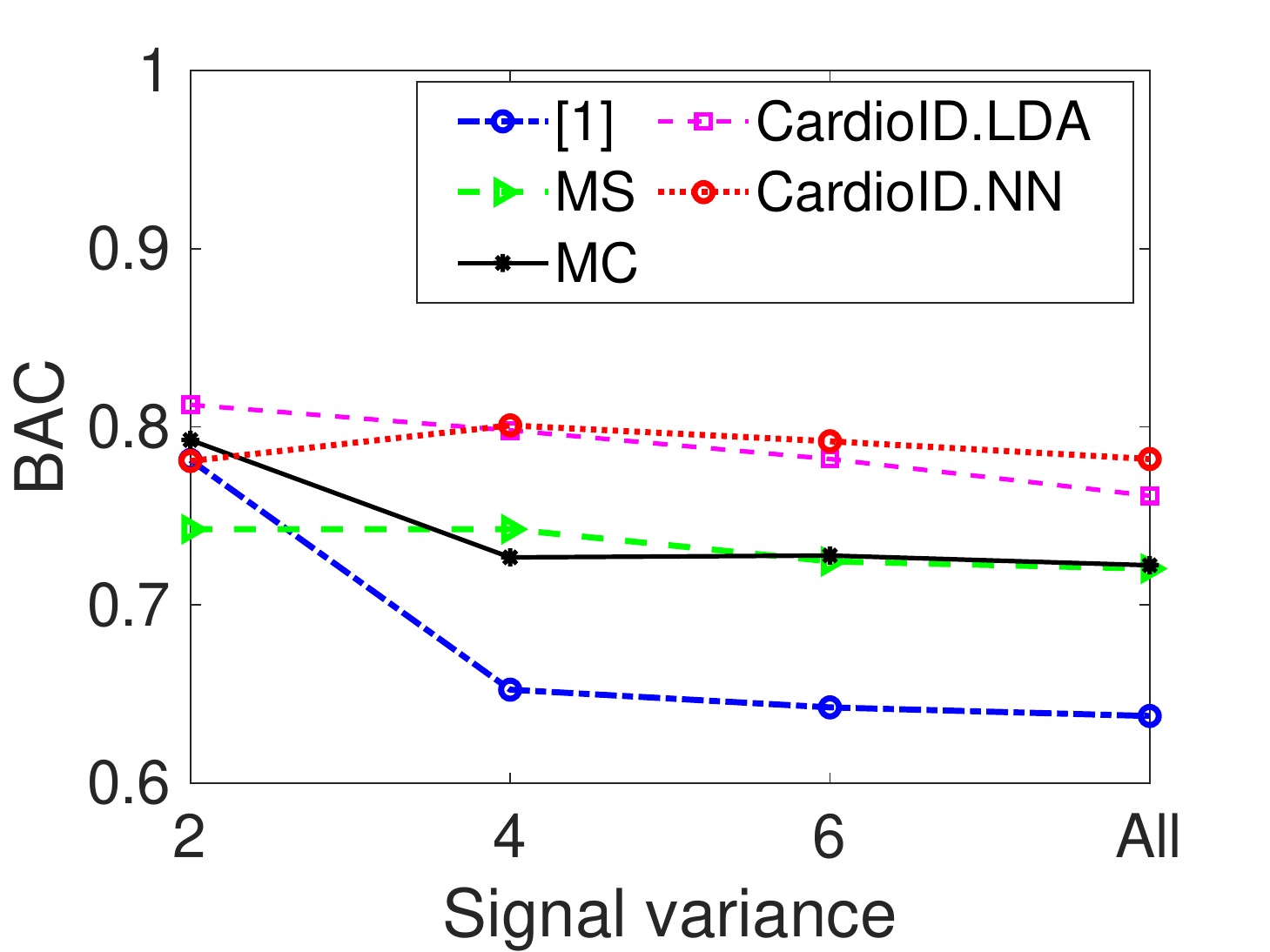}
      \vspace{-5mm}
      \caption{Camera dataset}
    \label{fig:camera_iden}
    \end{subfigure}
    \vspace{-1mm}
    \caption{Identification performance.}
    \label{fig:iden_performance}
    \vspace{-2mm}
\end{figure}

\textbf{Public dataset.} \autoref{fig:online_iden} shows the results for the pulse oximeter dataset. The MS variant provides the most significant improvement, above 10\% for most signal variances. This occurs because our harmonic filter adapts to every user, enabling more distinct and stable features. The MC variant does not really improve the BAC, but the fact that it accepts multiple morphologies improves inclusion (more subjects are accepted) and the acquisition rate (40\% higher than the MS variant and 20\% higher than the SoA baseline for the entire dataset~\cite{kavsaouglu2014novel}, cf. last column in \autoref{Tab:online_periods}). As for the \textit{LDA} and \textit{NN} variants, the performances are similar. The improvement of both solutions over the MC variant is around 5\%. The final BAC of \cardioid reaches 91.6\%, close to the 95\% target, while the SoA drops to 76.1\%~\cite{kavsaouglu2014novel}, leading to a total improvement of around 15\%.

\textbf{Private dataset.} \autoref{fig:camera_iden} shows the results for our camera dataset. We can see that, compared to the prior dataset, the overall performance is lower, but \textit{the same general trends appear, showcasing the ability of \cardioid to improve the identification performance across different sensors and users}. Considering the subsets with variances four and above, we obtain the same benefits as for the pulse oximeter:~the~MS variant provides about a 4\% improvement, the MC variant offers a marginal improvement in terms of accuracy but significant improvements on subject inclusion and period acquisition rate (cf. \autoref{Tab:camera_periods}), and the classification methods (LDA and NN) provide around a 7\% improvement.

It is important to note that even though \cardioid performs significantly better than the SoA with the camera dataset (12\% better), it is still far from the 95\% BAC target. In expectation, an 80\% BAC is not reliable because it gives the right user an 80\% chance to access the system and attackers a 20\% chance. At the end of this section we discuss some ways to overcome this problem.


\begin{figure*}[t!]
    \centering
    \begin{minipage}{.24\textwidth}
      \centering
      \vspace{-8mm}
      \includegraphics[width=\linewidth]{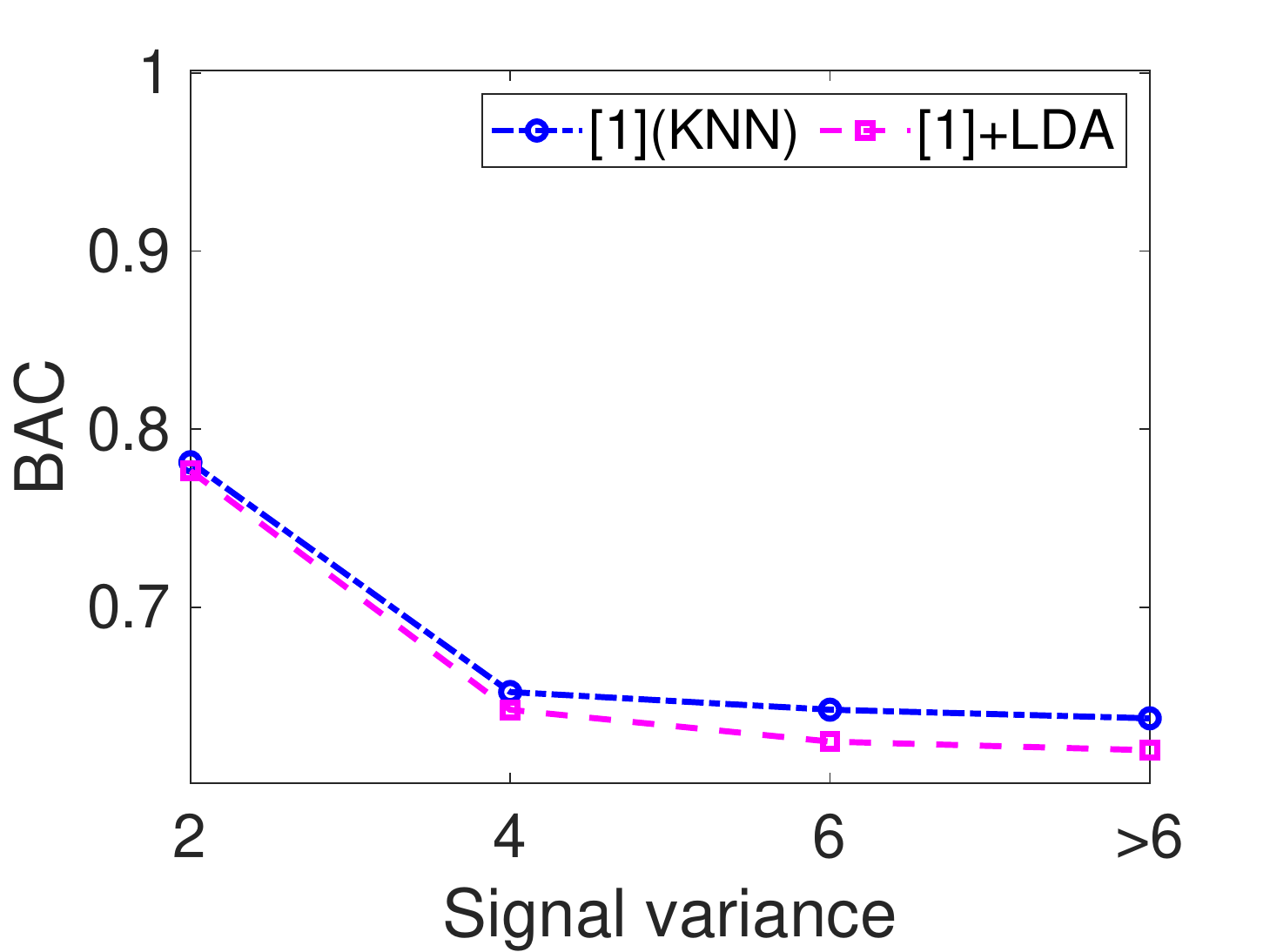}
      
      \caption{LDA contribution.}
      \label{fig:LDA_improve}
    \end{minipage}
    \begin{minipage}{.5\textwidth}
      \begin{subfigure}{.495\textwidth}
          \centering
          \vspace{-3mm}
          \includegraphics[width=\linewidth]{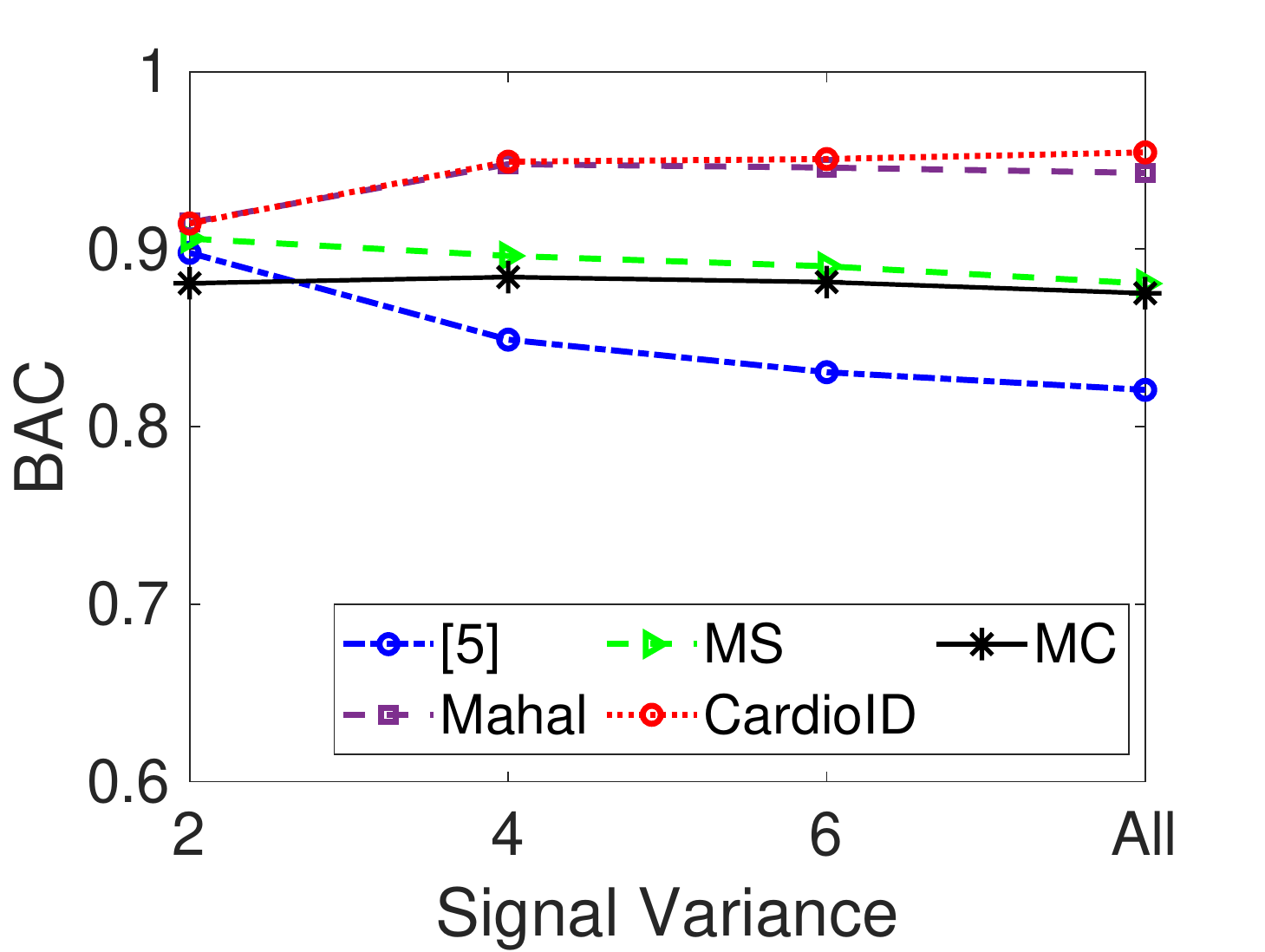}
          \caption{Pulse oximeter dataset.}
        \label{fig:online_authen}
        \end{subfigure}
        \begin{subfigure}{.495\textwidth}
          \centering
          \vspace{-3mm}
          \includegraphics[width=\linewidth]{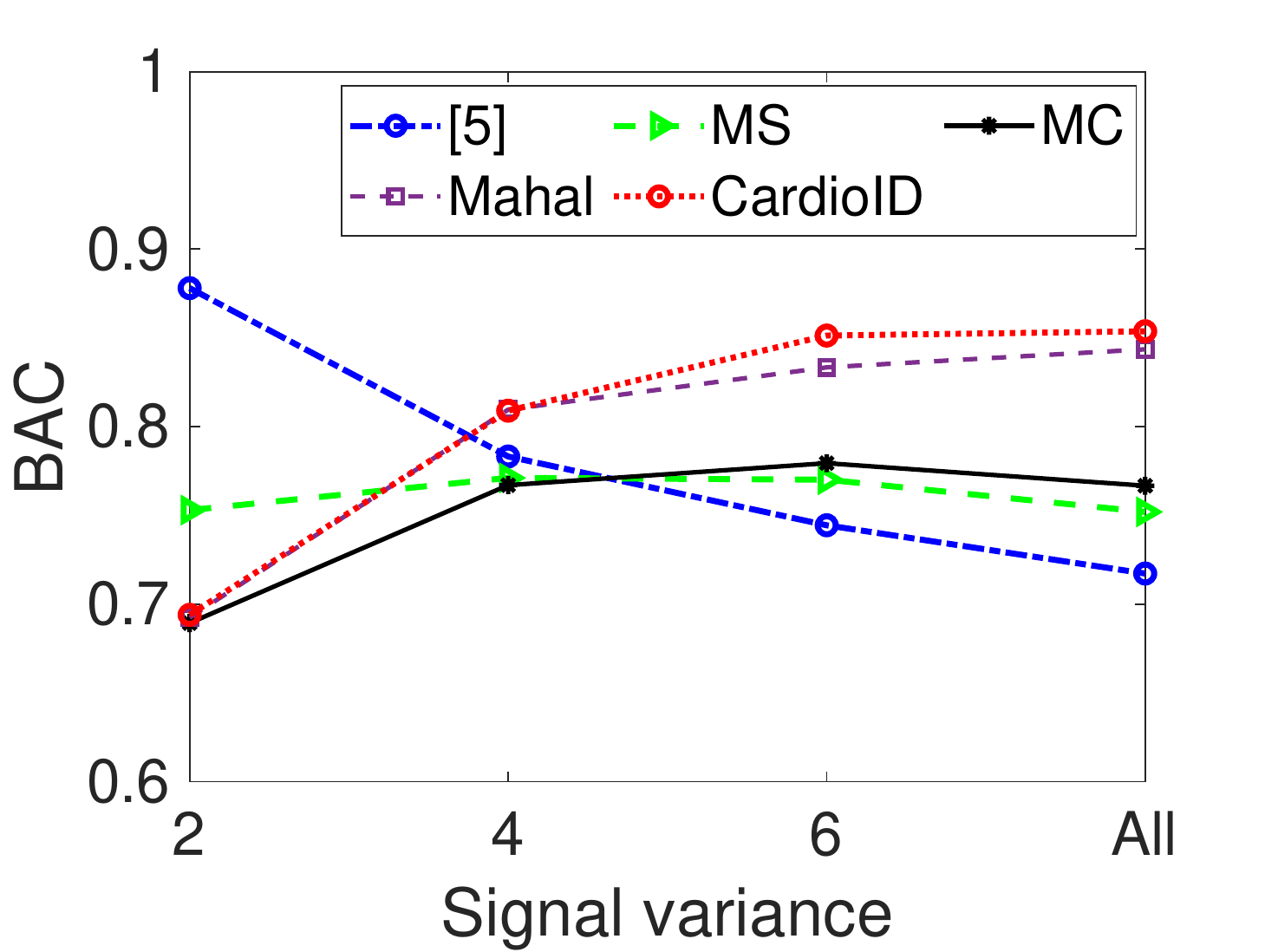}
          \caption{Camera dataset.}
        \label{fig:camera_authen}
        \end{subfigure}
        \caption{Authentication performance.}
        \label{fig:authen_performance}
    \end{minipage}
    \begin{minipage}{.24\textwidth}
      \centering
      \vspace{-6mm}
      \includegraphics[width=\linewidth]{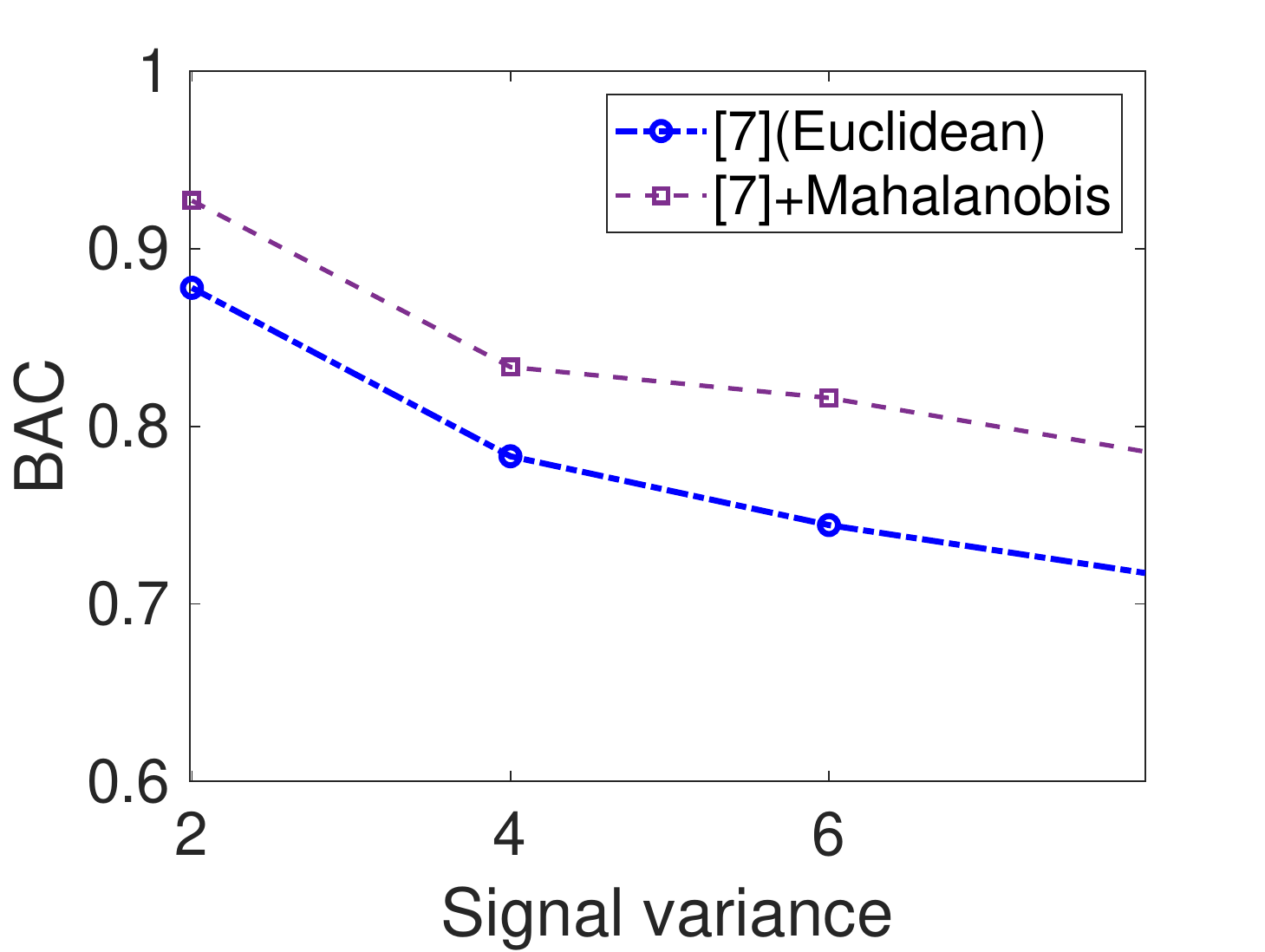}
      \caption{Mahal. contribution.}
      \label{fig:Mahal_improve}
    \end{minipage}
    \vspace{-5mm}
\end{figure*}

\textbf{Linear vs. non-linear methods.} \label{subsec:DR_reason}
SoA studies have been using linear~\cite{yadav2018emotion,bonissi2013preliminary,sarkar2016biometric} and non-linear~\cite{karimian2017human,spachos2011feasibility} methods to perform identification, but no study has compared both approaches or stated why one is preferable over the other. Our evaluation shows that both methods have a similar performance (\autoref{fig:iden_performance}). We hypothesize that this occurs because our morphology stabilization and classification provide cardiac periods with stable and distinct features, and hence, the role of the classification method is less prominent.  
To highlight the importance of our morphology variants (MS and MC), we replace the K-NN classifier used in~\cite{kavsaouglu2014novel}~with LDA in the camera dataset (while maintaining everything else the same).
The results in \autoref{fig:LDA_improve} show that LDA even degrades the performance a bit. Without the stable and distinct PPG signals provided by our morphology stabilization and classification, a machine learning method cannot do much on its own to overcome the high variance present in uncontrolled scenarios.


\subsection{Authentication} 

\textbf{Public dataset.}  
Before discussing the authentication results, we need to highlight a critical difference compared to identification: for authentication, the SoA~\cite{liu2019cardiocam} \textbf{fails} to include all types of users. In~\autoref{Tab:online_periods}, we can observe that both, the baseline for identification~\cite{kavsaouglu2014novel} and the MC variant, consider all 35 subjects for $t=4$ and above, and thus, the comparison is unbiased because the population size is the same. However, \textit{the maximum number of subjects considered by~\cite{liu2019cardiocam} (baseline for authentication), is only 29 (82.9\%)}. This occurs because our evaluation requires at least 20 periods per subject, but the morphological requirements of~\cite{liu2019cardiocam} are so stringent that we cannot get enough periods for 6 subjects. Hence, in authentication, \cardioid faces a more challenging scenario than the SoA because a bigger population increases the likelihood of errors. 

With that clarification, we can now discuss the main insights for the pulse oximeter dataset in \autoref{fig:online_authen}. First, at $t=2$, all the approaches have a similar performance. This occurs due to the limited data. For $t=2$, the emulated {subset} filters out most samples. Contrary to identification, where the system can exploit the samples from \textit{all the other users}, in authentication, the system can only use the limited samples belonging to a \textit{single} user. Thus, for $t=2$, the performance of the system is largely determined by the small number of relatively well controlled signals, leaving little room for the methods to showcase their respective strengths. For $t=4$ and above, however, we can observe that the MS and MC variants play the same role as in identification: MS increases the performance, while MC increases the participation (number of subjects) and the acquisition rate (reduces delay). Overall, \cardioid achieves a 93.7\% BAC with 35 subjects, while \cite{liu2019cardiocam} achieves 11\% less BAC with only 29 subjects.

\textbf{Private dataset.}
\autoref{fig:camera_authen} shows the results with the camera dataset. Due to the higher signal variability of this dataset, the SoA baseline filters even more periods than with the prior dataset: the percentage of subjects for the SoA is less in \autoref{Tab:camera_periods} than in \autoref{Tab:online_periods} (69.8\% vs. 82.9\%), and as stated before, three subjects did not have a single period satisfying the morphological requirements in~\cite{liu2019cardiocam}, which would make the SoA system futile for this target group. Even if we leave that critical point aside, CardioID still outperforms the SoA by 10\%, but it is still not able to reach the desired 95\% BAC target. A counter-intuitive trend with the camera dataset is that \cardioid's performance increases significantly with the signal variance. But this is not due to the increase in variance per se (which adds noise), but due to the increase in data.

There is one result, however, that we did not expect and it highlights a particular strength of the SoA with the camera: for $t=2$, ~\cite{liu2019cardiocam} has a strong performance compared to all of our variants, and it is in accordance with what the authors report in the original paper.\footnote{In~\cite{liu2019cardiocam}, the authors report a 95\% BAC using a single sample. Their signals seem to have a variance of 1.5. If we extrapolate the performance of~\cite{liu2019cardiocam} in \autoref{fig:camera_authen}, we will obtain the reported result.} 
Initially, we thought that it was because, at $t=2$, the SoA considers only {11} out of {43} subjects (25.6\% in \autoref{Tab:camera_periods}), but our MS variant, which relies only on morphology-2, also considers {the same amount of subjects} and performs worse. The SoA has stringent morphological requirements that filter out too many cardiac periods when the signal variability is high, but this conservative standard allows them to have more similarity among their periods when the data is more controlled, and it is particularly useful in authentication because the training phase utilizes a single user. The stability of our features relies solely on the harmonic filter. After that, the conditions to consider a morphology valid are rather permissive, simply counting the number of peaks and valleys present.

\begin{table*}[t!]
    \centering
    
    \caption{Most relevant studies in the SoA. Some studies evaluate multiple datasets, if the dataset is public, a reference is given in the `\# subjects' column. The studies in bold are used as comparison baselines.}
    \label{tab:summary_soa}
    \renewcommand{\arraystretch}{1.2}
    \resizebox{\linewidth}{!}{
    \begin{tabular}{|c|c|c|c|c|c|c|c|}
        \hline
         & \textbf{Application} & \textbf{Sensor Type} & \textbf{\# Subjects} & \textbf{\# Features} & \textbf{Decision Method} & \textbf{\# PT} & \textbf{Accuracy} \\
         \hline
         \cite{bonissi2013preliminary}, 2013 & Ident. & pulse oximeter & 44,14   & 1 (entire period) & correlation functions & 16 & 5.3\%,14.5\% (EER)\\
         \hline
         \textbf{\cite{kavsaouglu2014novel}, 2014} & Ident. & pulse oximeter & 30 & 40 (fiducial) & KNN & 1 & 87.0\% (F1) \\
         \hline
         \cite{sarkar2016biometric}, 2016 & Ident. & pulse oximeter & 32 , \cite{koelstra2011deap} & 15 (fiducial) & LDA \& QDA &  5 & 92.5\% (Rank-1)\\
         \hline
         \textbf{\cite{liu2019cardiocam}, 2019} & Auth. & camera & 25 & 30 (fiducial) & PCA & 1 & 95.8\% (BAC)\\
         &&&&\& 36 (non-fiducial)&&&\\
         \hline
         \cardioid & Both & Both & 35 , \cite{PPG2019dataset} &32,38,44 (fiducial) & LDA \& NN (Ident.)& 1& 93.2\% (Ident.), 95.5\% (Auth.) -- (BAC)\\
         &&& 43 && PCA (Auth.)&& 78.2\% (Ident.), 85.3\% (Auth.) -- (BAC)\\
         \hline
    \end{tabular}
    }
\end{table*}

\textbf{Mahalanobis contribution.}
A final point to discuss is the role of the Mahalanobis distance. Among all our variants, the Mahalanobis variant provides the biggest improvement,{\footnote{We also tried the variant MC+MultiCluster, without including the Mahalanobis distance, but the result of that variant was lower than the MC+Mahalanobis variant.}} which leads to the following question: If we simply replace the Euclidean distance with the Mahalanobis distance in \protect{\cite{liu2019cardiocam}} in the camera dataset, would the performance get comparable to \protect{\cardioid}? 
The result is shown in \autoref{fig:Mahal_improve}, where we can see that there is an improvement across all variances but the gain is not as significant as when Mahalanobis runs on top of MS and MC. When all signals are considered, the BAC of modified SoA (\cite{liu2019cardiocam}+Mahalanobis) is below {80\%}, while that of \protect{\cardioid} is {around 85\%}. Similar to what happened with identification, where we simply modify the SoA to bypass MS and MC (cf. \autoref{fig:LDA_improve}), this result proves that all the foundational blocks of \protect{\cardioid} are important to obtain a maximum performance gain.


\textbf{Summary.}
Based on the results with all datasets, we can summarize the following takeaway lessons. Overall, morphology stabilization (\autoref{sec:stabilization}) and the methods to overcome the non-linear effects of authentication (\autoref{sec:iden_authen}) improve performance, while morphology classification (\autoref{sec:feature}) improves the acquisition rate (more users and less latency). For the public dataset with uncontrolled data, our methods can bring back the accuracy of identification and authentication close to the desired target of 95\% BAC using a single cardiac period for testing. With the private and MIMIC-III dataset, however, even though our methods still have a better performance than the SoA, they do not reach the desired 95\% target. To ameliorate this problem, one could place stricter constraints on the types of morphologies that can be accepted and use more periods for testing, but that would increase the training overhead of the system and the delay while testing.

\section{Related Work}
\label{sec:soa}

We divide the related work into three main phases, highlighting the elements we build upon from the SoA and the novelty of our work. A summary of the most relevant studies is presented in~\autoref{tab:summary_soa}.

\textit{Phase 1: Basic identification.} Gu \textit{et. al.} report the first results for PPG identification using just four features. They achieve an accuracy of 94\% using a discriminant function~\cite{gu2003firstppg}.
Later, researchers found that the derivatives of a PPG signal can provide more stable and unique features~\cite{yao2007ppgderivatives}. Motivated by those initial results, researchers performed further experiments and found that the reported high accuracy is strongly dependent on the data gathering process. Spachos~\textit{et al.}~\cite{spachos2011feasibility} considers fiducial features and derivatives with two data sets. They report widely different performance for each set, EER (Equal Error Rate) 0.5\% vs. 25\% , leading them to state that PPG signals can be used for identification \textbf{\textit{``given that [they] are collected under controlled environments and with accurate sensors"}}. Bonissi~\textit{et al.}~\cite{bonissi2013preliminary} also find significant differences in EER depending on the databases they use, 5.3\% vs. 14.5\%. 

One of the most comprehensive evaluations is performed by Kavsao{\u{g}}lu~\textit{et al.}~\cite{kavsaouglu2014novel}. They use 40 features from PPG signals and their derivatives (first and second), and utilize a \textit{single period} for testing to obtain an accuracy of 87.2\% F1 score, which is stricter than BAC. From that work, we borrow the idea of using the second derivative and multiple features. We implemented this method and used it as a baseline of comparison for identification. 

\textit{Phase 2: non-fiducial features and more challenging PPG signals.} An important motivation for our study comes from~\cite{sarkar2016biometric} and~\cite{yadav2018emotion}. Sarkar~\textit{et al.}~\cite{sarkar2016biometric} analyze PPG signals with subjects that undergo various emotions. To enforce the same morphology, they normalize signals in time and amplitude so fiducial features can maintain a common pattern. This approach, however, requires 20 cardiac periods per testing sample ($\sim$15 sec, too long of a delay).  Yadav~\textit{et al.}~\cite{yadav2018emotion} also look into PPG signals that consider different levels of emotions and physical exercise, but they propose to use \textit{non}-fiducial features based on continuous wavelet transform (CWT). CWT considers the spectral response of a signal, which is more resilient to noise than geometric (fiducial) features, but they still require long testing sequences, between 8 and 40 cardiac periods ($\sim$6 to 60 sec). 

From the above studies, we take two insights. First, the normalization of PPG signals in time and amplitude to overcome distortions caused by emotions. Second, we do not use non-fiducial (spectral) features due to the many cardiac periods needed for testing, but we do perform a thorough spectral analysis (harmonic filtering) to obtain more stable and distinct fiducial features.


\textit{Phase 3: cameras and authentication.} Most studies focus on performing \textit{identification} with \textit{pulse oximeters}, but a recent work has been able to use smartphone \textit{cameras} to attain \textit{authentication}~\cite{liu2019cardiocam} (CardioCam). Authentication is more challenging than identification because it trains the system with a single user. CardioCam achieves a high BAC (95.8\%) using a single cardiac period for testing. 
We implement the \textit{signal processing chain} of CardioCam (filters, features and PCA method) and show that its performance decreases significantly with irregular PPG signals. Furthermore, for some users, the requirements of the CardioCam's morphology are so strict that they cannot use the system. 

There are also \textit{some other studies related to our work.} 

\textit{Cardiac health applications}. Several cardiac health applications use a smartphone camera. 
Chandrasekaran ~\textit{et al.}~\cite{chandrasekaran2012cuffless} combines sound information from the chest and video from a fingertip to measure people's blood pressure. Their estimation accuracy is above 95\%. 
HemaApp~\cite{wang2016hemaapp} infers hemoglobin levels based on the light absorption detected by a smartphone camera, and achieves sensitivity and precision of 85.7\% and 76.5\%. 
Despite of the distinct research purposes, we all need to tune cameras to obtain cardiac signals. 

\textit{Identification with ECG sensors.} The most common cardiac sensor is ECG. These sensors are widely available in hospitals and have also been used for identification. 
Safie~\textit{et al.} uses pulse active ratio to extract ECG features, and obtain an AUR and EER of 0.9101 and 0.1813, respectively \cite{safie2011electrocardiogram}.
Silva~\textit{et al.} explore a less invasive form of ECG sensors for user authentication, finger-ECG~\cite{da2013finger}. They utilize pre-processed templates as inputs for K-NN and SVM and obtain an EER below 9.1\%.
ECG sensors are more accurate than PPG sensors and smartphone cameras, but they are less pervasive and their filtering and identification methods are similar to the SoA studies with pulse oximeters.

\section{Conclusions}

Motivated by the fact that the study of PPG biometrics has been largely limited to controlled setups, we analyze the impact of more realistic \textit{uncontrolled} signals. We identify three main limitations in the SoA: the same filtering parameters are used to obtain the features for all users, a single dominant morphology is assumed, and there are important non-linear effects that have not been considered. Our solution, named \cardioid, overcomes those limitations with a novel morphology stabilization and classification mechanism, and by using the Mahalanobis distances with a multi-cluster approach. On average, with uncontrolled signals, \cardioid provides a BAC of 90\%, while the SoA attains a much lower BAC of 75\%. 

\balance

\bibliographystyle{abbrv}
\bibliography{main.bbl}

\begin{thebibliography}{10}

\bibitem{agrawal1998CLIQUE}
R.~Agrawal et~al.
\newblock Automatic subspace clustering of high dimensional data for data
  mining applications.
\newblock In {\em SIGMOD}, 1998.

\bibitem{ankerst1999optics}
M.~Ankerst et~al.
\newblock Optics: ordering points to identify the clustering structure.
\newblock {\em ACM Sigmod record}, 28(2):49--60, 1999.

\bibitem{YoshuaBengio}
Y.~Bengio.
\newblock How large should be the data set for training a deep auto encoder?,
  2011.
\newblock
  https://www.quora.com/How-large-should-be-the-data-set-for-training-a-Deep-auto-encoder.

\bibitem{bhattacharyya2009biometric}
D.~Bhattacharyya et~al.
\newblock Biometric authentication: A review.
\newblock {\em IJUNESST}, 2009.

\bibitem{bonissi2013preliminary}
A.~Bonissi et~al.
\newblock A preliminary study on continuous authentication methods for
  photoplethysmographic biometrics.
\newblock In {\em 2013 BIOMS}, 2013.

\bibitem{butterworth1930theory}
S.~Butterworth et~al.
\newblock On the theory of filter amplifiers.
\newblock {\em Wireless Engineer}, 1930.

\bibitem{carlsson2004total}
M.~Carlsson et~al.
\newblock Total heart volume variation throughout the cardiac cycle in humans.
\newblock {\em AJP Heart and Circulatory Physiology}, 2004.

\bibitem{chandrasekaran2012cuffless}
V.~Chandrasekaran et~al.
\newblock Cuffless differential blood pressure estimation using smart phones.
\newblock {\em IEEE TBME}, 2012.

\bibitem{ppgpressure2020}
A.~Chandrasekhar et~al.
\newblock Ppg sensor contact pressure should be taken into account for
  cuff-less blood pressure measurement.
\newblock {\em IEEE TBME}, 2020.

\bibitem{choi2017PPG}
A.~Choi and H.~Shin.
\newblock Photoplethysmography sampling frequency: pilot assessment of how low
  can we go to analyze pulse rate variability with reliability?
\newblock {\em Physiological measurement}, 2017.

\bibitem{da2013finger}
H.~P. Da~Silva, A.~Fred, A.~Louren{\c{c}}o, and A.~K. Jain.
\newblock Finger ecg signal for user authentication: Usability and performance.
\newblock In {\em BTAS}, 2013.

\bibitem{ester1996DBSCAN}
M.~Ester et~al.
\newblock A density-based algorithm for discovering clusters in large spatial
  databases with noise.
\newblock In {\em Kdd}, 1996.

\bibitem{Farooq2010segment}
U.~{Farooq} et~al.
\newblock Ppg delineator for real-time ubiquitous applications.
\newblock In {\em EMBC}, 2010.

\bibitem{gu2003firstppg}
Y.~Y. {Gu}, Y.~{Zhang}, and Y.~T. {Zhang}.
\newblock A novel biometric approach in human verification by
  photoplethysmographic signals.
\newblock In {\em ITAB}, 2003.

\bibitem{hall2015guyton}
J.~E. Hall.
\newblock {\em Guyton and Hall textbook of medical physiology e-Book}.
\newblock Elsevier Health Sciences, 2015.

\bibitem{hern2017samsung}
A.~Hern.
\newblock Samsung galaxy s8 iris scanner fooled by german hackers.
\newblock {\em The Guardian}, 23, 2017.

\bibitem{hinton2006autoencoder}
G.~E. Hinton and R.~R. Salakhutdinov.
\newblock Reducing the dimensionality of data with neural networks.
\newblock {\em science}, 313(5786):504--507, 2006.

\bibitem{israel2005ecg}
S.~A. Israel, J.~M. Irvine, A.~Cheng, M.~D. Wiederhold, and B.~K. Wiederhold.
\newblock Ecg to identify individuals.
\newblock {\em Pattern recognition}, 2005.

\bibitem{karimian2017human}
N.~Karimian et~al.
\newblock Human recognition from photoplethysmography (ppg) based on
  non-fiducial features.
\newblock In {\em ICASSP}, 2017.

\bibitem{kavsaouglu2014novel}
A.~R. Kavsao{\u{g}}lu et~al.
\newblock A novel feature ranking algorithm for biometric recognition with ppg
  signals.
\newblock {\em Computers in biology and medicine}, 2014.

\bibitem{koelstra2011deap}
S.~Koelstra et~al.
\newblock Deap: A database for emotion analysis; using physiological signals.
\newblock {\em IEEE transactions on affective computing}, 2011.

\bibitem{lee1999NMF}
D.~D. Lee and H.~S. Seung.
\newblock Learning the parts of objects by non-negative matrix factorization.
\newblock {\em Nature}, 401(6755):788--791, 1999.

\bibitem{lin2017cardiac}
F.~Lin et~al.
\newblock Cardiac scan: A non-contact and continuous heart-based user
  authentication system.
\newblock In {\em MobiCom}, 2017.

\bibitem{liu2019cardiocam}
J.~Liu et~al.
\newblock Cardiocam: Leveraging camera on mobile devices to verify users while
  their heart is pumping.
\newblock In {\em MobiSys}, 2019.

\bibitem{maaten2008tSNE}
L.~v.~d. Maaten and G.~Hinton.
\newblock Visualizing data using t-sne.
\newblock {\em Journal of machine learning research}, 9(Nov):2579--2605, 2008.

\bibitem{macqueen1967kmean}
J.~MacQueen et~al.
\newblock Some methods for classification and analysis of multivariate
  observations.
\newblock In {\em Proceedings of the fifth Berkeley symposium on mathematical
  statistics and probability}, 1967.

\bibitem{mahalanobis1936generalized}
P.~C. Mahalanobis.
\newblock On the generalized distance in statistics.
\newblock In {\em National Institute of Science of India}, 1936.

\bibitem{pearson1901pca}
K.~Pearson.
\newblock Liii. on lines and planes of closest fit to systems of points in
  space.
\newblock {\em The London, Edinburgh, and Dublin Philosophical Magazine and
  Journal of Science}, 1901.

\bibitem{peng2018BIRCH}
K.~Peng et~al.
\newblock Balanced iterative reducing and clustering using hierarchies with
  principal component analysis (pbirch) for intrusion detection over big data
  in mobile cloud environment.
\newblock In {\em SpaCCS}, 2018.

\bibitem{roweis2000LLE}
S.~T. Roweis and L.~K. Saul.
\newblock Nonlinear dimensionality reduction by locally linear embedding.
\newblock {\em science}, 290(5500):2323--2326, 2000.

\bibitem{safie2011electrocardiogram}
S.~I. Safie and et~al.
\newblock Electrocardiogram (ecg) biometric authentication using pulse active
  ratio (par).
\newblock {\em TIFS}, 2011.

\bibitem{sarkar2016biometric}
A.~Sarkar and et~al.
\newblock Biometric authentication using photoplethysmography signals.
\newblock In {\em BTAS}, 2016.

\bibitem{sheik2000wavecluster}
G.~Sheikholeslami et~al.
\newblock Wavecluster: a wavelet-based clustering approach for spatial data in
  very large databases.
\newblock {\em The VLDB Journal}, 2000.

\bibitem{PPG2019dataset}
A.~Siam and et~al.
\newblock Real-world ppg dataset.
\newblock Mendeley data, 2019.
\newblock http://dx.doi.org/10.17632/yynb8t9x3d.1.

\bibitem{spachos2011feasibility}
P.~Spachos and et~al.
\newblock Feasibility study of photoplethysmographic signals for biometric
  identification.
\newblock In {\em ICDSP}, 2011.

\bibitem{tenenbaum2000isomap}
J.~B. Tenenbaum et~al.
\newblock A global geometric framework for nonlinear dimensionality reduction.
\newblock {\em science}, 2000.

\bibitem{wang2016hemaapp}
E.~J. Wang and et~al.
\newblock Hemaapp: noninvasive blood screening of hemoglobin using smartphone
  cameras.
\newblock In {\em Ubicomp}, 2016.

\bibitem{Athlete2018heartrate}
G.~Whitworth.
\newblock Why do athletes have a lower resting heart rate?, 2018.
\newblock https://www.healthline.com/health/athlete-heart-rate.

\bibitem{yadav2018emotion}
U.~Yadav and et~al.
\newblock Evaluation of ppg biometrics for authentication in different states.
\newblock In {\em ICB}, 2018.

\bibitem{yao2007ppgderivatives}
J.~{Yao} and et~al.
\newblock A pilot study on using derivatives of photoplethysmographic signals
  as a biometric identifier.
\newblock In {\em EMBC}, 2007.

\bibitem{zhao2013human}
Z.~Zhao et~al.
\newblock A human ecg identification system based on ensemble empirical mode
  decomposition.
\newblock {\em Sensors}, 2013.

\end{thebibliography}

\end{document}